\documentclass[12pt,twoside]{book}
\usepackage[T1]{fontenc}
\usepackage[Sonny]{fncychap}
\usepackage[b5paper,centering]{geometry}
\usepackage[backend=biber,style=numeric,sorting=none,maxnames=10]{biblatex}
\usepackage[acronym,nomain,nonumberlist]{glossaries}
\usepackage{glossary-mcols}
\usepackage[spanish,activeacute,es-tabla]{babel}
\usepackage[autostyle, spanish = mexican]{csquotes}
\usepackage{epigraph}
\usepackage{graphicx}
\usepackage{bmpsize}
\usepackage{multicol}
\usepackage{multirow}
\usepackage{amssymb}
\usepackage{colortbl}
\usepackage{xcolor}
\usepackage{framed}
\usepackage{tabularx}
\usepackage{booktabs}
\usepackage{caption}
\usepackage{float}
\usepackage{fancyhdr}
\usepackage{url}
\usepackage{datetime}
\usepackage{xifthen}
\usepackage{tcolorbox}
\usepackage{tikz}
\usepackage{xparse}
\usepackage{empheq}
\usepackage{afterpage}
\usepackage[pagecolor=white,nopagecolor=white]{pagecolor}
\usepackage{covington,tocloft}
\usepackage{titletoc}
\graphicspath{{images/}}

\makeglossaries
\usetikzlibrary{matrix,arrows, positioning,shadows,shadings,backgrounds,calc,shapes, tikzmark}
\tcbuselibrary{skins,breakable,listings,theorems}
\ChNameVar{\Large\sf\color{titlepagecolor}}
\ChNumVar{\Huge\color{titlepagecolor}}
\ChTitleVar{\large\sf\color{titlepagecolor}}  
\ChRuleWidth{0.5pt}
\ChNameUpperCase
\definecolor{row1}{rgb}{0.95,0.95,0.95}
\definecolor{row2}{rgb}{0.85,0.85,0.85}
\definecolor{titlepagecolor}{cmyk}{1,.60,0,.40}
\definecolor{namecolor}{cmyk}{1,.50,0,.10}
\definecolor{azulF}{cmyk}{1,.60,0,.40} 
\definecolor{verdep}{cmyk}{1,.50,0,.10}

\newenvironment{declaration}[1] {
\pagestyle{empty}
  \begin{center}
  \vspace*{1cm}
  {\Large \bfseries #1}
  \end{center}
  \vspace{0.5cm}
   \begin{quote}}
{\end{quote}
}

\contentsmargin{0cm}
\titlecontents{chapter}[0pc]
{\addvspace{30pt}%
  \begin{tikzpicture}[remember picture, overlay]%
  \draw[fill=verdep!70,draw=verdep!70] (-4,-.1) rectangle (-0.1,0.5);%
  \pgftext[left,x=-1.5 cm,y=0.2cm]{\color{azulF}\Huge\sc\bfseries \thecontentslabel};%
  \end{tikzpicture}\color{azulF}\large\sc\bfseries%
}%
{}
{}
{\;\color{verdep!70}\titlerule\;\color{azulF}\large\sc\bfseries \thecontentspage
\begin{tikzpicture}[remember picture, overlay]
\draw[fill=verdep!70,draw=verdep!70] (1pt,0) rectangle (6,0.1pt);
\end{tikzpicture}}%

\titlecontents{section}[2.4pc]
{\addvspace{1pt}}
{\contentslabel[\color{azulF}\thecontentslabel]{2.4pc}}
{}
{\hfill\small \color{azulF}\thecontentspage}
[]
\titlecontents*{subsection}[4pc]
{\addvspace{-1pt}\small}
{}
{}
{\hfill\small \color{azulF}\thecontentspage}
[\\][]

\makeatletter
\renewcommand{\tableofcontents}{%
\chapter*{%
\vspace*{-20\p@}%
\begin{tikzpicture}[remember picture, overlay]%
\pgftext[right,x=15cm,y=0.2cm]{\color{azulF}\Huge\sc\bfseries \contentsname};%
\draw[fill=verdep!70,draw=verdep!70] (13,-1) rectangle (20,1.75);%
\clip (13,-1) rectangle (20,1.75);
\pgftext[right,x=15cm,y=0.2cm]{\color{azulF}\Huge\sc\bfseries \contentsname};%
\end{tikzpicture}}%
\@starttoc{toc}}
\makeatother

\makeatletter
\newlength \figwidth
\if@twocolumn
\else
\fi
\makeatother

\newcommand{\InsertFig}[7]{%
	\ifthenelse{\isempty{#5}}%
	{% if #5 is empty
		\begin{figure}[htbp!]
		{\ifthenelse{\isempty{#7}}{\vspace{0mm}}{\vspace{#7}}}
		\centering
		\includegraphics[#6]{#1}%
		\caption{#3}{\textbf{#4}}
		\label{#2}
		\end{figure}    
	}
	{% if #5 is not empty
		\begin{figure}[htbp!]
		{\ifthenelse{\isempty{#7}}{\vspace{0mm}}{\vspace{#7}}}
		\centering
		\includegraphics[width=#5\textwidth,#6]{#1}%
		\caption{#3}{\textbf{#4}}
		\label{#2}
		\end{figure}
	}
}

\newcommand{\listofexamples}{Índice de ejemplos}
\newlistof{example}{lex}{\listofexamples}
\newcommand{\lentry}{\addcontentsline{lex}{example}{\protect\numberline{Ejemplo
\thetcbteo}}}

\definecolor{colordominanteD}{RGB}{74,0,148}
\newcounter{tcbteo}[chapter] % Contador
\renewcommand{\thetcbteo}{\thechapter.\arabic{tcbteo}}%Formato 1.1,1.2,..
\tikzset{nodoTeorema/.style={
    rectangle, top color=gray!5, bottom color=gray!5,
    inner sep=1mm,anchor=west,font=\small\bf\sffamily}
}
\newtcolorbox{cajaTeorema}[3][]{
    arc=0mm,breakable,enhanced,colback=gray!5,boxrule=0pt,top=7mm,
    drop fuzzy shadow, fontupper=\normalsize,
    step and label={tcbteo}{#3},
overlay unbroken= {
    \draw[colordominanteD,line width =2.5cm]([xshift=1.25cm, yshift=0cm]frame.north west)--+(0pt,3pt);
    \draw[color=colordominanteD,line width =0.2pt](frame.north west)--([xshift=0pt]frame.north east);
    \node[nodoTeorema](tituloteo)
        at ([xshift=0.2cm, yshift=-4mm]frame.north west){\textbf{\color{colordominanteD} Ejemplo \thetcbteo\lentry \;#2}};
},
overlay first = {
    \draw[colordominanteD,line width =2.5cm]([xshift=1.25cm, yshift=0cm]frame.north west)--+(0pt,3pt);
    \draw[color=colordominanteD,line width =0.2pt](frame.north west)--([xshift=0pt]frame.north east);
    \node[nodoTeorema](tituloteo)
        at ([xshift=0.2cm, yshift=-4mm]frame.north west){\textbf{\color{colordominanteD} Ejemplo \thetcbteo\lentry \;#2}};
}, %First
overlay middle = { },
overlay last = { }
#1}

\NewDocumentEnvironment{ejemplo}{O{} O{} O{}}{
    \bigskip\begin{cajaTeorema}{#1}{#2}
    #3
}{\end{cajaTeorema}\bigskip }

\newacronym{ared}{ARED}{Adaptive Random Early Detection}
\newacronym{wred}{WRED}{Weighted Random Early Detection}
\newacronym{wfq}{WFQ}{Weighted Fair Queuing}
\newacronym{tcp}{TCP}{Transport Control Protocol}
\newacronym{udp}{UDP}{User Datagram Protocol}
\newacronym{qos}{QoS}{Quality of Service}
\newacronym{ip}{IP}{Internet Protocol}
\newacronym{ietf}{IETF}{Internet Engineering Task Force}
\newacronym{isa}{ISA}{Integrated Service Architecture}
\newacronym{tos}{ToS}{Type of Service}
\newacronym{tc}{TC}{Traffic Class}
\newacronym{isp}{ISP}{Internet Service Provider}
\newacronym{sla}{SLA}{Service Level Agreement}
\newacronym{owd}{OWD}{One-Way Delay}
\newacronym{rtt}{RTT}{Round-Trip Time}
\newacronym{voip}{VoIP}{Voice over Internet Protocol}
\newacronym{mos}{MOS}{Mean Opinion Score}
\newacronym{itu}{ITU}{International Telecommunication Union}
\newacronym{fps}{FPS}{First Person Shooter}
\newacronym{cac}{CAC}{Call Admission Control}
\newacronym{ftp}{FTP}{File Transfer Protocol}
\newacronym{p2p}{P$2$P}{Peer-to-Peer}
\newacronym{vps}{VPS}{Variable Packet Size}
\newacronym{pptd}{PPTD}{Packet Pair/Train Dispersion}
\newacronym{slops}{SLoPS}{Self-Loading Periodic Streams}
\newacronym{topp}{TOPP}{\textit{Trains Of Packet Pairs}}
\newacronym{abett}{ABETT}{Available Bandwidth Estimations Techniques and Tools}
\newacronym{pgm}{PGM}{Probe Gap Model}
\newacronym{prm}{PRM}{Probe Rate Model}
\newacronym{adsl}{ADSL}{Asymmetric Digital Subscriber Line}
\newacronym{ixp}{IXP}{Internet eXchange Point}
\newacronym{bdp}{BDP}{Bandwidth Delay Product}
\newacronym{ns}{NS}{Network Simulator}
\newacronym{cbq}{CBQ}{Class Based Queueing}
\newacronym{pq}{PQ}{Priority Queueing}
\newacronym{fifo}{FIFO}{First In First Out}
\newacronym{aqm}{AQM}{Active Queue Management}
\newacronym{red}{RED}{Random Early Detection}
\newacronym{p2ptv}{P2P-TV}{Peer-to-Peer Television}
\newacronym{pymes}{PYMES}{Pequeñas y Medianas Empresas}
\newacronym{sip}{SIP}{Session Initiation Protocol}
\newacronym{iax}{IAX}{Inter-Asterisk eXchange protocol}
\newacronym{toip}{ToIP}{Telephony over IP}
\newacronym{rtp}{RTP}{Real-time Transport Protocol}
\newacronym{http}{HTTP}{Hypertext Transfer Protocol}
\newacronym{jpeg}{JPEG}{Joint Photographic Experts Group}
\newacronym{pal}{PAL}{Phase Alternating Line}
\newacronym{ntsc}{NTSC}{National Television System Committee}
\newacronym{dsl}{DSL}{Digital Subscriber Line}
\newacronym{svc}{SVC}{Scalable Video Coding}
\newacronym{avl}{AVL}{Adaptative Video Layering}
\newacronym{nat}{NAT}{Network Address Translation}
\newacronym{qoe}{QoE}{Quality of Experience}
\newacronym{mpegts}{MPEG-TS}{Moving Picture Experts Group - Transport Stream}
\newacronym{ts}{TS}{Transport Stream}
\newacronym{mtu}{MTU}{Maximum Transfer Unit}
\newacronym{iptv}{IPTV}{Internet Protocol Television}
\newacronym{cbr}{CBR}{Constant Bit Rate}

\glsaddall

\begin{document}

% \maketitle

% ---------------------------------------------------
% Portada
% ----------------------------------------------------------------
\begin{titlepage}
\newgeometry{left=7.3cm} %defines the geometry for the titlepage
\newpagecolor{titlepagecolor}\afterpage{\nopagecolor}
\noindent
\color{white}
\makebox[0pt][l]{\rule{1.3\textwidth}{2pt}}
\vspace*{1mm}
\par
\noindent
\textbf{{\huge \textsf{El impacto del \textit{buffer} en la calidad de\\ servicio}}}% \textcolor{namecolor}{\textsf{Heidelberg}}
\vfill
\noindent
% {\huge \textsf{Luis Sequeira}}
\textsf{Luis Sequeira}
% \vskip\baselineskip
% \noindent
% \textsf{Febrero 2021}
\end{titlepage}
\restoregeometry % restores the geometry
% \nopagecolor % Use this to restore the color pages to white

% ---------------------------------------------------
% Ficha catalográfica
% ---------------------------------------------------
\thispagestyle{empty}
% \hrule
\begin{center}
    \begin{tabular}{m{1.5cm}m{9cm}}
        \hline
        004 &  \\
        S773 -i & Sequeira Villarreal, Luis Enrique\\
         & \hspace*{0.5cm}El impacto del \textit{buffer} en la calidad de servicio Luis Enrique Sequeira Villarreal-- 1a. ed. \\
         & -- San José: Luis Sequeira, 2021.\\
         & \hspace*{0.5cm}1  recurso en línea ; 1.6Mb.\\
         & \\
         & \hspace*{0.5cm}ISBN  978-9968-49-653-7\\
         & \\
         & \hspace*{0.5cm}1. Redes  -- 2. Comunicaciones informáticas\\
         & \hspace*{1.5cm}-- I. Título.\\
        \hline
    \end{tabular}
\end{center}
% \hrule

% ---------------------------------------------------
% Copyright
% ---------------------------------------------------
\vspace*{\fill}
\noindent
\copyright \ Luis Sequeira \the\year \\
\hspace*{3mm} Primera edición \\
% \vspace*{\fill}
% \begin{center}
%     \begin{pspicture}[shift=*](3.5,3)
%         \psbarcode{978-9968-49-653-7}{includetext}{isbn}
%     \end{pspicture}
% \end{center}
\cleardoublepage

% correct bad hyphenation here
\hyphenation{op-tical net-works semi-conduc-tor dis-po-si-ti-vos dis-po-si-ti-vo di-fe-ren-tes re-pre-sen-ta-ti-vo res-pon-sa-ble res-pon-sa-bles co-rres-pon-dien-te co-rres-pon-dien-tes ad-mi-nis-tra-do la-bo-ra-to-rio acce-so buffer}

% The frontmatter text starts here
\frontmatter
\pagestyle{fancy}
\fancyhf{}

\cleardoublepage
\begin{declaration}{Prefacio}
En este texto se analiza la respuesta de la transmisión de flujos de datos en tiempo real, en escenarios de redes de acceso, en los cuales dichos flujos convergen en un enlace de salida, compitiendo por alcanzar un determinado nivel de calidad de servicio. La concurrencia de este tipo de flujos puede generar ráfagas de paquetes, que en determinadas circunstancias pueden comprometer la capacidad que tienen los \textit{buffer} para absorber paquetes en períodos de congestión.

Además, se presenta un análisis de las características de los \textit{buffer} en los dispositivos de acceso, especialmente su tamaño y la pérdida de paquetes. En particular, se describe cómo estas características pueden afectar a la calidad de las aplicaciones multimedia cuando estas generan tráfico a ráfagas y sus posibles efectos en el tráfico de otras aplicaciones que comparten un enlace en común.

El contenido es una versión revisada y editada de partes de la tesis doctoral ``\textit{Técnicas de estimación de \textit{buffer}, centradas en las redes de acceso, para la transmisión de flujos IP en tiempo real}'' \cite{ls18} publicada en Zaragoza, España, $2015$.

Se agradece a Idelkys Quintana por las revisiones detenidas del libro, señalando errores en el texto, sugiriendo mejoras y por todos sus valiosos comentarios.\\

\vspace*{0.5cm}

Londres, Febrero \the\year. \hfill Luis Sequeira
\end{declaration}

\cleardoublepage

\tableofcontents 

\printglossary[title=Acrónimos,style=long,type=\acronymtype]

\mainmatter

\pagestyle{fancy}
\fancyhf{}
\fancyhead[LE,RO]{\leftmark}
\fancyhead[RE,LO]{}
\fancyfoot[CE,CO]{\thepage}
\fancyfoot[LE,RO]{}

\chapter{Introducción}
\label{cha:Introduccion}
El desarrollo de aplicaciones o el despliegue de nuevos servicios que envían datos a red, requiere de una planificación desde el punto de vista de red, ya que esto permitirá garantizar un determinado nivel de calidad, lo cual repercutirá en el grado de satisfacción de los usuarios finales. En este ámbito, resulta curioso cómo profesionales de disciplinas diferentes, o incluso de ámbitos similares, pueden abordar un problema de maneras muy diferentes o identificar potenciales riesgos en soluciones viables. Por ejemplo, en el caso de una aplicación para videoconferencia, el desarrollador de software podría enfocarse en proveer la más alta calidad de video posible, para así obtener un mayor impacto en el mercado e incrementar el número de usuarios satisfechos con una interfaz gráfica de vanguardia y una experiencia de alta calidad en el video. Al mismo tiempo, un profesional de redes podría señalar un potencial riesgo en el deterioro de la calidad debido a la gran cantidad de datos que se debe transmitir en tiempo real desde un usuario a otro utilizando una red como Internet. En una red descentralizada como esta, no se tiene control de la configuración óptima de los dispositivos de red intermedios entre los usuarios finales, y por lo tanto, no se pueden emplear ciertas técnicas de ingeniería de tráfico en los equipos que conforman el camino de red entre dos nodos. 

Una posible solución para satisfacer los diferentes puntos de vista descritos anteriormente, podría ser desarrollar la aplicación de manera que se pueda adaptar a diferentes condiciones de la red, las cuales podrían cambiar incluso durante una misma sesión de usuario. La calidad variaría en función de la capacidad o de las limitaciones de la red. La máxima calidad en términos de video se presentaría bajo condiciones óptimas de la red. Por el contrario, la calidad de video bajaría cuando la red presente peores condiciones. Claramente no es el caso ideal, pero el usuario no perdería la sesión y podría seguir utilizando el servicio.

El ejemplo anterior no es un caso aislado si se tiene en cuenta el amplio crecimiento del número de usuarios y de los nuevos servicios multimedia e interactivos en Internet (por ejemplo: video bajo demanda, videoconferencia, juegos en línea, \gls{voip}, \textit{streaming}, videovigilancia, etc.). Estos servicios generan una cantidad significativa de tráfico en la red \cite{games3, camera1}. Además, la expectativa de crecimiento para estas aplicaciones, indica que la tendencia de uso se incrementará. Además, los usuarios de dichos servicios demandan mejores experiencias en el uso de las aplicaciones multimedia. Muchos de estos servicios derivan de aplicaciones en tiempo real desarrolladas sobre redes específicas como redes de conmutación de circuitos y tienen estrictos requerimientos de calidad. Sin embargo, las diversas tecnologías de acceso a Internet son bastante heterogéneas, dificultando en muchos casos, la provisión de servicios que satisfagan las expectativas de los usuarios. Este es el principal motivo por el cual es necesario tener en cuenta la \gls{qos} que estos servicios ofrecen para sus aplicaciones; especialmente cuando las tecnologías de acceso deben soportar aplicaciones y servicios multimedia en tiempo real.

En este ámbito, el presente libro ofrece una serie de análisis con ejemplos sencillos y prácticos, enfocados para que profesionales de diversas áreas puedan comprender el comportamiento del tráfico de diversos servicios y su impacto en la \gls{qos}, ya que tiene un efecto considerable en el consumo de recursos en las redes de acceso. Cada servicio genera un tráfico con características muy particulares, el cual varía en función de la naturaleza y el tamaño de la información. 

La calidad se describe desde el punto de vista de los \textit{buffer}, debido a que son utilizados como mecanismos de regulación de tráfico en los dispositivos de red. Además, cuando se discute la planificación de una red, el tamaño del \textit{buffer} de los nodos es un parámetro importante de diseño, ya que existe una relación entre dicho tamaño y la utilización del enlace. Cuando el \textit{buffer} está lleno y la cantidad de memoria es grande, generará un incremento significativo en la latencia, a este fenómeno se le conoce como \textit{bufferbloat}. Por otro lado, si la cantidad de memoria es muy pequeña, se incrementará la pérdida de paquetes en los nodos durante los períodos de congestión. Como consecuencia, la influencia del \textit{buffer} debería ser considerada cuando se trata de mejorar la utilización del enlace y la calidad de aplicaciones y servicios. Existen muchos estudios en relación al dimensionado de \textit{buffer} \cite{buffers5}, pero están especialmente enfocados a los \textit{router} del núcleo de la red y para flujos \gls{tcp}, trabajan sobre diferentes estructuras de colas y no contemplan en detalle el comportamiento diferente de los \textit{buffer}.

\section{Cuellos de botella}

Por otro lado, el crecimiento en la demanda de datos y las complejas arquitecturas de red que se presentan hoy en día, producen que ciertos puntos en la red, fuera de la red troncal, se conviertan en cuellos de botella. Esto sucede principalmente en las redes de acceso, ya que las capacidades son menores que en las redes de transporte; aunque estos puntos críticos de congestión también pueden presentarse en redes de altas prestaciones, incluso en la nube. En estos puntos, generalmente en el \textit{router} de acceso, la principal causa de pérdida de paquetes es el descarte en las colas. Es por esto, que la implementación del \textit{buffer} en los nodos de red y sus políticas de gestión son de gran importancia para asegurar la entrega del tráfico de las diferentes aplicaciones y servicios.

En un entorno residencial, donde el trabajar desde casa es ahora muy común (principalmente durante el inusual año 2020), o para las pequeñas empresas, los efectos del comportamiento del tráfico o de los \textit{buffer} de los \textit{router} pueden ser más pronunciados, debido a las modestas infraestructuras de acceso que estos puedan tener. Así, las características de diseño del \textit{buffer} del nodo de la red y las políticas de gestión que este implemente, tienen una gran importancia a la hora de asegurar la entrega correcta del tráfico de diferentes aplicaciones y servicios, por lo que, sería útil tener en cuenta los parámetros y el comportamiento del \textit{buffer} junto a la estimación de la capacidad del enlace.

Tradicionalmente, el ancho de banda disponible, el retardo y \textit{jitter} entre dos dispositivos finales de red, se han utilizado como parámetros que dan una idea general de la \gls{qos} que se podría tener en un determinado enlace. Pero, hoy en día, se sabe que estos parámetros pueden verse afectados por el comportamiento de los \textit{buffer} que se encuentran entre los equipos terminales \cite{ls4, ls8}. Dicho comportamiento está determinado principalmente por el tamaño y las políticas de gestión de los \textit{buffer} (es decir, la manera en que se llena y se vacía el \textit{buffer}). De tal manera, que la pérdida de paquetes puede ser causada por los \textit{buffer}, cuyo comportamiento a su vez, también podrían modificar ciertos parámetros de \gls{qos}. 

Es bien conocido que los \textit{router} del núcleo hacen un uso extensivo de diversas técnicas para la gestión de colas de manera activa o \gls{aqm}, las cuales son capaces de mantener la longitud de la cola más pequeña que las tradicionales \textit{drop-tail}, lo cual previene el \textit{bufferbloat} y reduce la latencia. En esta área hay algoritmos muy conocidos como \gls{red} y algunos derivados \gls{ared} o \gls{wred}, pero estos algoritmos requieren de un ajuste cuidadoso de sus parámetros con el fin de proveer un buen rendimiento \cite{buffers13}. También, existen algoritmos de planificación de la \gls{qos} como \gls{wfq}, el cual es una técnica de planificación de paquetes de datos, que permite establecer estadísticamente, una serie de prioridades a flujos multiplexados. Sin embargo, estas soluciones presentadas en la mayor parte de los estudios de investigación se aplican sobre estructuras de colas y no son aplicables a las redes de acceso que normalmente utilizan \textit{router} de gama media y baja, los cuales no suelen implementar técnicas avanzadas de gestión de tráfico, e incluso, en la mayoría de los equipos sólo hay un \textit{buffer} tipo \gls{fifo} \cite{buffers11}.

Por otra parte, es cierto que lo más ampliamente estudiado ha sido el rendimiento de \gls{tcp} y que una gran cantidad de variantes se han desplegado (por ejemplo, SACK, New Reno, Vegas, etc.) con el fin de mejorar determinadas características adaptándose a las diferentes situaciones de la red. No obstante, muchas aplicaciones multimedia y servicios en tiempo real transportan su información sobre \gls{udp}, de tal manera, que las aplicaciones tienen que ser capaces de descubrir el comportamiento de la red para poder optimizar el tráfico.

Muchos servicios y aplicaciones multimedia que se transportan sobre \gls{udp} (por ejemplo, videoconferencia, video \textit{streaming}, \gls{voip}, entre otros) utilizan herramientas que permiten la estimación del ancho de banda disponible, conocidas como \gls{abett} \cite{bw1, bw2}, para mejorar la utilización del enlace y algunos parámetros de QoS. Todas estas herramientas tienen dos cosas en común: se enfocan en las estimaciones de los enlaces que conforman el núcleo de la red y no tienen en cuenta el comportamiento del \textit{buffer} y sus parámetros.

\section{Tráfico a ráfagas}

Es habitual que algunos servicios generen tráfico a ráfagas, como por ejemplo los sistemas de videovigilancia, videoconferencia, \textit{streaming} de video, IPTV y otros servicios interactivos. Este comportamiento se presenta cuando se debe enviar una gran cantidad de información (\textit{frames} de video o imágenes) en un tiempo muy corto. Dichas ráfagas pueden incluir diferentes números de paquetes, y eventualmente, podrían congestionar ciertos dispositivos de red cuando la cantidad de paquetes transmitidos es significativa con respecto al tamaño del \textit{buffer}. En dicha situación, existen aplicaciones que son desarrolladas utilizando herramientas de conformado de tráfico para proveer cierta \gls{qos} y una mejor experiencia al usuario, y al mismo tiempo no ser tan perjudicial para la red, pero incrementando el consumo de recursos del dispositivo debido al procesamiento.

También se debe considerar que algunos de estos servicios y aplicaciones de Internet, generan paquetes cuyos tamaños pueden variar desde unas pocas decenas de $ bytes $ (como ocurre en el caso de \gls{voip}) hasta otros que utilizan tamaños mayores (por ejemplo, videoconferencia o videovigilancia). Sin embargo, el tamaño del \textit{buffer} y el ancho de banda disponible, pueden estar dimensionados para que dichos parámetros se mantengan estables, creando problemas de congestión en enlaces de acceso sensibles.

\section{Alcances}

Como se ha mencionado anteriormente, este libro pretende introducir al lector, con una serie de conceptos básicos y ejemplos, en el rol que juega el \textit{buffer} de los dispositivos de red en la \gls{qos}. Lo cual es un aspecto a tener en cuenta a la hora de realizar el desarrollo de aplicaciones, despliegue de servicios, la planificación de una red o cuando se quiere proveer ciertos niveles de \gls{qos}. 

El capítulo \ref{cha:Calidad de Servicio} presenta brevemente las definiciones más comunes de los parámetros \gls{qos}, como el retardo, \textit{jitter} y la pérdida de paquetes; además describe algunas medidas objetivas y subjetivas de la calidad. Finalmente se comentan algunas métricas y técnicas de estimación del ancho de banda disponible. El capítulo \ref{cha:Buffer} describe los principales conceptos de los \textit{buffer}, donde se definen varias técnicas de dimensionado y disciplinas de colas. Además, se habla del efecto del desbordamiento y la influencia en diferentes servicios. También, se analiza el caso del posible desbordamiento de los \textit{buffer} cuando el enlace tiene un bajo nivel de utilización y la influencia que dichos \textit{buffer} tienen en la \gls{qos} de diferentes servicios. Por otro lado, se han seleccionado los servicios \gls{voip}, videovigilancia, videoconferencia, video \textit{streaming} y \gls{p2ptv}, a modo de ejemplo, para analizar los aspectos más importantes que definen la manera en que dichos servicios envían tráfico a la red; esto se comenta en el capítulo \ref{cha:Servicios Multimedia}.

Los capítulos \ref{cha:Buffer y rafagas} y \ref{cha:Buffer y applicaciones} tienen un enfoque diferente, un poco más práctico a modo de ejemplos. Se han definido dos casos de uso de flujos \gls{ip} en tiempo real que pueden considerarse comúnmente utilizados a nivel residencial, pequeñas empresas, ciertos modelos de negocios y grupos de usuarios. El objetivo es analizar la respuesta en la transmisión de los flujos \gls{ip} en tiempo real cuando estos comparten un enlace con servicios que generan tráfico a ráfagas y sus repercusiones en la \gls{qos}, debido al efecto del \textit{buffer}. Además, valorar el aumento de la capacidad de la red interna como posible solución bajo estas condiciones. Finalmente, el capítulo \ref{cha:Conclusion} resume los aspectos más relevantes presentados en este libro.

\chapter{La Calidad de Servicio}
\label{cha:Calidad de Servicio}
% \minitoc
La redes \gls{ip} fueron diseñadas en un contexto donde las aplicaciones eran relativamente tolerantes a los retardos, a las posibles pérdidas de paquetes, y a los enlaces de modesta capacidad y con baja demanda de tráfico \cite{stallings}, como por ejemplo el tráfico de sitios web de Internet y servicios como \gls{ftp}. Sin embargo, en los últimos años, estas redes se han desplegado ampliamente por todo el mundo, dando paso a una gran cantidad de nuevos tipos de servicios con requerimientos diferentes, así como a un aumento importante en la cantidad de usuarios a nivel global. Muchos de los servicios que se utilizan en la actualidad, son en tiempo real o tienen requerimientos de cierta interactividad, como sucede en los juegos en línea, por mencionar un ejemplo. En general, estos servicios son muy sensibles a los retardos, ya que en algunos casos, puede que no tenga sentido procesar un paquete de datos si el retardo es muy grande, como sucede en muchos servicios interactivos. 

La congestión es un factor que puede comprometer la calidad de un servicio, debido a que los \textit{buffer} de los nodos de la red pueden descartar paquetes que no pueden procesar en un momento dado, y por lo tanto, generar pérdida de paquetes, degradando la calidad de los servicios. Para hacer frente a esta situación no basta con incrementar la capacidad de una red o en algunos casos no es posible. En dichos casos, se hace necesario implementar mecanismos que gestionen el tráfico y controlen la congestión. Para satisfacer estas necesidades, la \gls{ietf} ha desarrollado un conjunto de estándares bajo el marco general de \gls{isa} \cite{rfc1633}. Los servicios integrados o también llamados \textit{IntServ} pretenden gestionar los recursos necesarios para garantizar la \gls{qos}, realizando una reserva extremo a extremo de recursos en los elementos que conforman la red.

La \gls{ietf} también ha desarrollado otra serie de estándares denominados servicios diferenciados o \textit{DiffServ} que proporcionan un método que busca garantizar la calidad de servicio \cite{rfc2475} de una manera más simple y de bajo coste. El modelo de servicios diferenciados analiza flujos de datos en vez de reservas de recursos. Para realizar un tratamiento diferenciado de la calidad de servicio, los paquetes \gls{ip} son etiquetados utilizando el campo \gls{tos} de la cabecera IPv4 \cite{ipv4} o \gls{tc} en IPv6 \cite{ipv6}. Esto significa que la negociación será realizada para todo el tráfico de una red, ya sea un \gls{isp} o una empresa. A dichas negociaciones se les llama \gls{sla}. Este tipo de acuerdos especifican qué clases de tráfico serán provistos y qué garantías se darán a cada flujo de tráfico.

Sin embargo, independientemente de la manera de gestionar la \gls{qos}, los parámetros de red suelen ser los mismos y el efecto que estos provocan varía en función del tipo de servicio ya que algunos toleran en cierta medida la pérdida de paquetes y otros son muy sensibles al retardo y el \textit{jitter}.

\section{Parámetros objetivos de QoS}

La percepción que tienen los usuarios de la \gls{qos} de un servicio en tiempo real (por ejemplo, la voz o el video), está relacionada con ciertos parámetros objetivos de la red, como lo son el retardo, el \textit{jitter} y la pérdida de paquetes.

\subsection{Retardo}

El retardo es un parámetro crítico para ciertas aplicaciones y servicios en tiempo real (por ejemplo en aplicaciones de robótica en la nube, la telecirugía \cite{ls7} o incluso los vehículos conectados \cite{ls10, ls9}); es también un factor de importancia en el diseño de las redes \cite{ls16, ls17} y en ocasiones es utilizado como una medida del rendimiento de una red. Puede definirse como el tiempo que necesita un dato para viajar a través de la red desde un nodo a otro. El retardo depende de muchos factores; entre los cuales se pueden mencionar \cite{park}:

\begin{itemize}
    \item El sistema de codificación.
    \item La paquetización.
    \item La codificación del canal.
    \item El retardo del paquete en los \textit{buffer}.
    \item El retardo de propagación.
\end{itemize}

En redes de paquetes es común utilizar el término \gls{owd} para referirse al retardo en un sentido de la comunicación (de fuente a destino), además se utiliza \gls{rtt} para designar al tiempo de ida y vuelta de un paquete. 

El retardo puede tener un impacto importante en la calidad percibida por un usuario, ya que en niveles altos puede ocasionar problemas de comunicación para servicios interactivos. Por ejemplo, si un teléfono \gls{ip} o \textit{gateway} \gls{voip} se conecta a través de una línea de baja velocidad donde el retardo puede ser significativo, se puede percibir eco o incluso problemas de interacción en la comunicación. Otro caso más crítico es el de la telecirugía donde los dispositivos hápticos necesitan una respuesta en términos de latencia de menos de $1 \; ms$ \cite{ls7}, lo cual es muy bajo.

\subsection{\textit{Jitter}}

Usualmente, los servicios en tiempo real requieren que los tiempos de llegada entre los paquetes sean constantes, de tal manera que puedan ser reproducidos por la aplicación en el tiempo correspondiente, como sucede con el tráfico de \gls{voip} o de videoconferencia. Sin embargo, las redes introducen retardos a los paquetes que difieren en su magnitud. Esta fluctuación de la magnitud del retardo se denomina \textit{jitter} y se define como la variación máxima de retardo que experimentan los paquetes en una sola sesión \cite{stallings}. Uno de los principales factores que causan el \textit{jitter} es la variación del retardo en los \textit{buffer} de los nodos de la red.

Para mitigar el efecto del \textit{jitter}, algunas aplicaciones introducen un \textit{buffer}, también llamado \textit{de-jitter buffer} \cite{gtc25}, el cual tiene la función de retardar ligeramente los paquetes para poder entregarlos a una velocidad constante al software que genera la señal de salida, por ejemplo audio o video. El \textit{jitter} es un factor crítico para algunas aplicaciones en tiempo real, ya que cuanto mayor sea la variación del retardo que estas permitan, más grande será el retardo real en la entrega de los datos, y por lo tanto, mayor será el tamaño del \textit{buffer} de \textit{de-jitter} que se necesitará en la recepción. 

\subsection{Pérdida de paquetes}

La pérdida de paquetes es uno de los principales problemas a los que se enfrentan las redes de comunicaciones. Se dice que hay pérdida de paquetes cuando un paquete no ha podido llegar a su destino. Este problema se puede presentar por diversos motivos, algunos de ellos relacionados con la degradación de la señal en el medio de comunicación, problemas en el \textit{hardware} o \textit{driver}. También se da el caso en el que los paquetes son descartados por políticas específicas de gestión de tráfico, con la intención de mantener cierto rendimiento de la red. Un ejemplo de esto es el descarte de paquetes en los \textit{buffer} de los nodos de la red, el cual puede darse en períodos de congestión. La pérdida de paquetes también puede darse por el \textit{buffer} de \textit{de-jitter}, ya que, cuando un paquete llega demasiado tarde para ser reproducido por la aplicación es desechado y para efectos prácticos es como si no hubiera llegado.

El protocolo \gls{tcp} posee mecanismos para solicitar la retransmisión de un paquete que no ha llegado a su destino \cite{tcp2}, pero esto conlleva un aumento del retardo en la red y no todos los servicios pueden ser capaces de tolerarlo. Además, es necesario tener en cuenta que una gran parte de los servicios en tiempo real utilizan \gls{udp} como protocolo de transporte.

\section{Medidas objetivas y subjetivas de QoS}

En las redes de telecomunicaciones, la calidad es uno de los aspectos de medición que tiene un alto nivel de importancia. Por lo tanto, la capacidad de una monitorización continua de la calidad es una prioridad para mantener la satisfacción de los usuarios de un determinado servicio (por ejemplo redes móviles o banda ancha). Por otro lado, muchas empresas implementan herramientas de monitorización (usualmente basadas en la nube) de la actividad de los usuarios en los servicios que proveen, la idea es mejorar la experiencia del usuario, ayudar a detectar y detener posibles amenazas de seguridad, entre otras funciones. 

El método más fiable de obtener una medida veraz de la percepción de un usuario con respecto a la calidad de un servicio es desarrollar adecuadamente un test basándose en ciertos criterios de satisfacción y aplicarlo a los usuarios para obtener una medida subjetiva de la calidad percibida por ellos \cite{voip1}. No obstante, este tipo de medición es lenta y su coste puede llegar a ser muy elevado, haciéndolo inadecuado para la monitorización en tiempo real.

Como alternativa, existen diversos modelos de medidas objetivas de la calidad, que proporcionan una evaluación automática de los sistemas de comunicación sin la necesidad de la intervención de los usuarios. Estas medidas objetivas utilizan modelos matemáticos para determinar los niveles de calidad que pueden ser fácilmente computarizados. Por lo general, se basan en la medida de los parámetros de \gls{qos} presentados en el apartado anterior (retardo, \textit{jitter} y pérdidas). Debido a la naturaleza heterogénea del tráfico de los distintos servicios que transporta una red, se puede decir, que es necesario un modelo diferente por cada tipo de servicio o aplicación.

En términos de voz, la calidad se refiere a la claridad con que una persona percibe la voz en una comunicación. La medición de la calidad de voz resulta, a menudo, muy útil en la evaluación de la gestión de los servicios de una red telefónica, como se muestra en el Ejemplo \ref{eje:mos} y el Ejemplo \ref{eje:emodel}.

\begin{ejemplo}[Medida subjetiva de calidad para voz.][eje:mos]

El \gls{mos} es una medida subjetiva que se utiliza cuando es necesario valorar los efectos subjetivos en la calidad de la voz, por ejemplo, cuando se incluye algún nuevo equipo de transmisión o se realizan modificaciones en las características de la transmisión de una red telefónica. Los métodos para obtener evaluaciones subjetivas de los sistemas y componentes de transmisión se encuentran estandarizados por la \gls{itu}. El \gls{mos} está definido en la recomendación P. $ 800 $ \cite{mos2}, en la cual cada percepción de los usuarios se clasifica en una escala subjetiva de $1$ a $5$, ver Figura \ref{fig:mos_levels}.

\begin{figure}[H]
    \centering
    \includegraphics{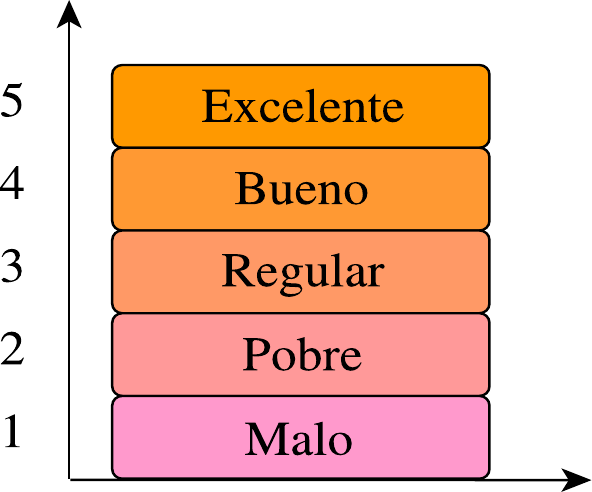}
    \caption{Escala de calidad para el \gls{mos} según ITU-T.}
    \label{fig:mos_levels}
\end{figure}

\end{ejemplo}

\begin{ejemplo}[Medida de calidad utilizando parámetros objetivos.][eje:emodel]

El \textit{E-model} ofrece una alternativa al \gls{mos} y consiste en otra recomendación de la \gls{itu}-T \cite{emodel}, la cual se usa como instrumento para estimar un posible nivel de calidad en función de distintos parámetros de \gls{qos} medidos de forma objetiva (retardo, pérdidas, etc.), proporcionando un medio para estimar el \gls{mos} \cite{mos1}. En general, se puede decir que el \textit{E-model} consiste en medir el \gls{mos} en un entorno controlando los parámetros de \gls{qos}, de tal forma que sepamos qué es lo que diría un usuario medio acerca de la calidad percibida en determinadas condiciones.

\end{ejemplo}

Los juegos en línea forman parte de otra área donde la calidad es muy importante, ya que los usuarios pueden tener una gran interactividad con el juego o con otros usuarios. En estos casos es necesario crear métricas de calidad personalizadas para cada juego en particular con la finalidad de obtener una idea de la calidad que experimentan los jugadores, ver Ejemplo \ref{eje:quake}.

\begin{ejemplo}[Medida de calidad para \textit{Quake IV}][eje:quake]

En \cite{games5}, los autores propusieron un método, para medir la calidad extremo a extremo, que permite cuantificar la calidad percibida de juegos \textit{online} interactivos. La metodología se expone utilizando un juego llamado \textit{Quake IV}, el cual es del tipo \gls{fps} y con gran aceptación a nivel mundial. \\

El método se denomina \textit{G-model} (por su similitud con el \textit{E-model}) y se llevó a cabo mediante una serie de experimentos subjetivos para cuantificar el impacto de los parámetros de la red, en la calidad percibida por los usuarios. Las pruebas se realizaron en una red \textit{Gigabit} con un servidor y $ 6 $ clientes con excelentes prestaciones de \textit{hardware} para gráficos con el fin de minimizar los posibles errores. Además, entre los enlaces de los clientes al servidor se introducen retardo, \textit{jitter} y pérdida de paquetes mediante Netem (\textit{Network Emulator}), el cual forma parte del kernel de Linux en las distribuciones actuales. Las pruebas realizadas demuestran que el \textit{G-model} permite predecir un \gls{mos} o calificación de calidad basando en valores medidos de retardo y \textit{jitter} con una correlación muy alta ($ R=0.98 $) con los datos subjetivos.

\end{ejemplo}

\section{Disponibilidad}

En términos generales, la disponibilidad se refiere a cuánto tiempo un dispositivo o sistema está operativo respecto al tiempo total que se hubiese deseado que funcionase. Por otro lado, es necesario relacionar el ancho de banda con la calidad obtenida en un determinado servicio y para ello se utiliza el concepto de disponibilidad. Se puede decir que cuanto mayor sea el ancho de banda más disponibilidad se puede tener.

En las redes de paquetes, el término ancho de banda a menudo se utiliza para caracterizar la cantidad de datos que una red puede transferir por unidad de tiempo. La estimación del ancho de banda es un parámetro de interés cuando se desea optimizar el rendimiento de transporte de extremo a extremo, por ejemplo, en el enrutamiento de una red o la distribución de contenidos en sistemas \gls{p2p}. Además, esta estimación es importante para el soporte de la ingeniería de tráfico y la planificación de la capacidad de la red. 

Existen varias métricas relacionadas con el ancho de banda (capacidad y ancho de banda disponible). En la actualidad, existen herramientas de estimación de ancho de banda que emplean diversas estrategias para medir estos parámetros. A lo largo de esta sección se presentan algunas de estas técnicas, así como, herramientas de medición de los parámetros mencionados.

\subsection{Métricas relacionadas con el ancho de banda}

La capacidad de un enlace se define como la cantidad máxima de información en $ bits $ que se puede enviar en un segundo. Es muy común que en un enlace (a nivel de capa $ 2 $) sea posible transmitir a una tasa de $ bit $ constante, la cual está limitada por la tecnología de la red, que marca el ancho de banda permitido en el medio de propagación y también por las limitaciones del \textit{hardware} en los dispositivos transmisores y receptores. Por ejemplo, esta tasa es de $ 10 \; Mbps $ para Ethernet $ 10BaseT $, de $ 1.544 \; Mbps $ para un $ T1 $ o de $ 2.048 \; Mbps $ para un $ E1 $. Sin embargo, para enlaces inalámbricos esto no es cierto, ya que algunas tecnologías de capa $ 2 $ no trabajan con tasas de transmisión constante \cite{ieee802.11}, como sucede en los sistemas inalámbricos IEEE $802$.$11$, los cuales conmutan la tasa en función de las características del medio y la tasa de error.

A nivel \gls{ip}, la capacidad que se pueden alcanzar es inferior debido al proceso de encapsulado. Para explicar este fenómeno, se supondrá un enlace con una capacidad a nivel de capa $ 2 $, $ C_{L_{2}} $, y un encabezado de capa $ 2 $, $ H_{L_{2}} $, entonces el tiempo $ t_{L_{3}} $ necesario para transmitir un paquete \gls{ip} de tamaño $ L_{L_{3}} $ es:

\begin{equation}
	t_{L_{3}}=\frac{L_{L_{3}}+H_{L_{3}}+H_{L_{2}}}{C_{L_{2}}}
\end{equation}

Por lo tanto, la capacidad en la capa $ 3 $, $ C_{L_{3}} $, es: 

\begin{eqnarray}
	\label{eq_capacidad}
	\nonumber C_{L_{3}} &=& \dfrac{L_{L_{3}}+H_{L_{3}}}{t_{L_{3}}} \\
	\nonumber           &=& \dfrac{L_{L_{3}}+H_{L_{3}}}{\dfrac{L_{L_{3}}+H_{L_{3}}+H_{L_{2}}}{C_{L_{2}}}} \\
	\nonumber           &=& C_{L_{2}}\dfrac{L_{L_{3}}+H_{L_{3}}}{L_{L_{3}}+H_{L_{3}}+H_{L_{2}}} \\
	           &=& C_{L_{2}}\dfrac{1}{1+\dfrac{H_{L_{2}}}{L_{L_{3}}+H_{L_{3}}}}
\end{eqnarray}

Nótese que la capacidad \gls{ip} descrita en la ecuación \ref{eq_capacidad} depende de la relación entre los tamaños del encabezado (de capa $ 2 $) y el paquete \gls{ip} (con su respectivo encabezado de capa $ 3 $). Como el uso del protocolo \gls{ip} está tan generalizado y con la finalidad de uniformizar el término independientemente de la tecnología, para efectos del presente libro, se va a definir la capacidad de extremo a extremo como la máxima tasa de transferencia posible medida a nivel \gls{ip}. Desde el punto de vista de un camino de red, la capacidad está limitada por el enlace con la mínima capacidad en dicho camino, a este enlace se le conoce como \textit{narrow link}. 

\begin{ejemplo}[La capacidad]
Si un paquete de $ 150 \; bytes $ (incluyendo la cabecera \gls{ip}) se transmite entre dos nodos adyacentes en un enlace Ethernet $ 10BaseT $ con una $ C_{L_{2}} $ de $ 10 \; Mbps $, el cual tiene un encabezado, $ H_{L_{2}} $, de $ 38 \; bytes $, la capacidad \gls{ip}, $ C_{L_{3}} $, es 

\begin{equation}
    \nonumber C_{L_{3}} = 10 \; Mbps \times \dfrac{1}{1+\dfrac{38 \; bytes}{150 \; bytes}} = 7.97 \; Mbps
\end{equation}

mientras que si el paquete tiene un tamaño de $ 1500 \; bytes $, la capacidad sería 

\begin{equation}
    \nonumber C_{L_{3}} = 10 \; Mbps \times \dfrac{1}{1+\dfrac{38 \; bytes}{1500 \; bytes}} = 9.75 \; Mbps
\end{equation}

\end{ejemplo}

Por otro lado, en una comunicación serial (como Ethernet, WiFi, etc.) solamente es posible enviar un \textit{bit} a la vez, el cual se envía por el enlace, a la máxima tasa que se puede alcanzar en un determinado instante, desde este punto de vista la utilización del enlace solamente tiene dos estados: ocupado o libre. Por este motivo, el ancho de banda disponible requiere ser definido en términos de una media de tiempo de la utilización instantánea, en una determinada transmisión o período de prueba.

El ancho de banda disponible es un término relacionado al ancho de banda que no se utiliza o que queda libre en un enlace durante un determinado período. Como se mencionó anteriormente, la capacidad de un enlace depende de las características de la capa de transmisión de una determinada tecnología y el medio de propagación. Sin embargo, el ancho de banda disponible está ligado a la carga de tráfico y su comportamiento en un determinado enlace, y usualmente es una métrica que varía en función del tiempo \cite{bandwidth4, gtc16} y el comportamiento de las aplicaciones que comparten el enlace.

Esta métrica puede ser influenciada por diversos factores, entre los que se destacan:  

\begin{itemize}
    \item El tráfico compuesto por la combinación de aplicaciones que utilizan \gls{tcp} y \gls{udp} cuando estas comparten un mismo enlace. 
    \item La implementación de los mecanismos de control de congestión que se incluyen en la recomendación del RFC $ 3782 $ \cite{rfc3148}, incluida cada variante de \gls{tcp} (como, Tahoe, Reno, New Reno \cite{rfc3782}, SACK \cite{rfc2018}, etc.) que permite alcanzar un nivel diferente de \textit{throughput}. 
    \item Aspectos como el tamaño de las tramas, el comportamiento y tamaño de los \textit{buffer} en los extremos de la red, la capacidad y la carga del enlace.
    \item El número de conexiones que compiten en un mismo enlace también influyen para el cálculo de ancho de banda disponible.
\end{itemize}

Por este motivo, las aplicaciones con determinados requerimientos de \gls{qos} usualmente deben adaptarse a las variaciones del ancho de banda disponible, y por lo tanto, necesitan medirlo con relativa rapidez ya que puede variar drásticamente en una misma sesión, incluso a lo largo del día en función de la carga de la red.

\subsection{Técnicas de estimación del ancho de banda}

El desarrollo de las técnicas de estimación de ancho de banda es un tema muy estudiado. En \cite{bandwidth2} y \cite{bandwidth3} se proponen las primeras herramientas para la estimación \gls{vps}. Además, se encuentra una gran cantidad de técnicas para la estimación del ancho de banda como \gls{pptd} \cite{bandwidth5, bandwidth6}, \gls{slops} \cite{bandwidth7} y \gls{topp} \cite{bandwidth8, bandwidth9}. Estas técnicas son conocidas como \gls{abett} y la mayoría de ellas se clasifican en dos grandes tendencias: \gls{pgm} y \gls{prm}.

Los métodos \gls{pgm} se caracterizan por ser rápidos y fáciles de implementar. Utilizan el muestreo de paquetes para observar la dispersión de los tiempos entre ellos y así estimar un ancho de banda disponible. Este tipo de técnicas tiene la desventaja que los resultados no son muy precisos en entornos con múltiples saltos \cite{bandwidth10}, además, asumen que sólo existe un cuello de botella de extremo a extremo. En \cite{bandwidth11}, se presenta una herramienta simple y ligera para la medición del ancho de banda disponible, sin embargo los autores afirman que necesita ser mejorada en relación con la precisión a la hora de realizar las estimaciones y que no se pueden realizar mediciones en el mismo equipo donde se está ejecutando.

Los \gls{prm} difieren de los \gls{pgm} en que son herramientas intrusivas, hacen uso de sondas de prueba que inducen un estado de congestión en la red para poder realizar medidas o estimaciones \cite{bandwidth12}. Las estimaciones se realizan enviando tráfico de prueba, si este se envía a una tasa menor que el ancho de banda disponible, la tasa de prueba corresponde a la tasa de salida en el otro extremo de la red, por el contrario si la tasa de prueba es mayor, los paquetes se encolan en los \textit{buffer} intermedios generando retardos y tasas de salida menores. Estos métodos presentan mayor precisión que los PGM pero el tiempo necesario para la estimación y la intrusión son sus principales desventajas. En \cite{bandwidth13}, se presenta una herramienta que funciona basada en los principios comentados anteriormente. Su principal ventaja es que se ejecuta en un sistema operativo en tiempo real, lo cual aumenta la estabilidad de las estimaciones y permite probar varias tasas con un solo flujo de paquetes.

\subsection{Medidas de disponibilidad}

Muchas veces es necesario medir o estimar la cantidad de conexiones que se pueden establecer sin pérdidas de datos, en un enlace con cierto ancho de banda disponible y para un servicio determinado. Esto es equivalente a medir o calcular el número de servidores disponibles en un determinado sistema, con lo que puede calcularse la probabilidad de bloqueo del sistema y utilizarse como parámetro de calidad relacionado con la disponibilidad del servicio. Con esta información las empresas pueden estimar el efecto de incorporar un nuevo servicio en la red teniendo en cuenta el número de servidores necesarios para dicho servicio y los efectos producidos por su tráfico correspondiente, o bien, valorar la posibilidad de un incremento del ancho de banda en un enlace, para poder disponer de un número de servidores tal que la disponibilidad del servicio sea aceptable. En estos casos, es necesario estimar cuánto tráfico requiere la disponibilidad de un nuevo servidor. Dicha estimación se realiza en función de la tecnología de la red, ya que depende de aspectos como el protocolo de acceso al medio, el tamaño de la trama y los encabezados utilizados por un determinado servicio. 

Se puede poner como ejemplo la telefonía \gls{ip} (ver Ejemplo \ref{eje:telefonia}) que es una de las soluciones utilizadas para mitigar los costes de la telefonía tradicional y donde la disponibilidad es fundamental. Sin embargo, en muchos casos las soluciones libres o propietarias carecen de mecanismos adecuados para proporcionar la \gls{qos} necesaria. Esto sucede porque no siempre se puede disponer de los servidores necesarios dado que a mayor número de servidores, mayor tráfico en la red, factor que puede disminuir la calidad. 

\begin{ejemplo}[Telefonía \gls{ip}][eje:telefonia]

En \cite{gtc1}, los autores estiman diferentes parámetros de calidad en la implementación de un sistema \gls{cac} en un entorno virtualizado. Los resultados sugieren que el aumento en el número de llamadas de voz repercute negativamente en la pérdida de paquetes de otras aplicaciones que comparten la red.\\

En \cite{gtc3}, se presenta un esquema de multiplexión de paquetes \gls{voip} para diferentes políticas de \textit{buffer}. La multiplexión consiste en incluir en un mismo paquete, los paquetes de diferentes flujos de llamadas \gls{ip}. Los autores afirman que dicho esquema reduce el ancho de banda, lo que permite más servidores, pero introduce nuevos retardos debido a la retención y al procesamiento en ambos extremos de la comunicación lo que disminuye la calidad.

\end{ejemplo}

En general, un análisis similar puede realizarse para otros servicios como la videoconferencia o los sistemas de videovigilancia, por mencionar algunos que se analizarán más adelante. Esto debido a que la probabilidad de bloqueo (y por lo tanto, la disponibilidad del servicio) es un parámetro que puede utilizarse indistintamente en diferentes servicios.

\chapter{El \textit{Buffer}}
\label{cha:Buffer}
% \minitoc
Internet puede definirse como un conjunto descentralizado de redes de comunicación, por esto, la arquitectura es bastante heterogénea. Los diferentes nodos en la red difieren en cuanto a su capacidad de procesamiento, memoria y ancho de banda. Además, las velocidades de entrada y salida de un \textit{router} pueden tener grandes diferencias dependiendo de la tecnología o los accesos utilizados, por ejemplo, las redes Ethernet tienen tasas de $ 10 $, $ 100 $ y $ 1000 \; Mbps $ (incluso más, pero no son comunes en redes de acceso), una red WiFi, puede tener una velocidad desde $ 2 $ hasta $ 54 \; Mbps $ para $ 802$.$11g $ o hasta $ 600 \; Mbps $ para IEEE $ 802$.$11n $, incluso pudiendo superar $ 1 \; Gbps $ para IEEE $ 802$.$11ac $. Por otro lado, las tecnologías asimétricas de acceso como cable módem y \gls{adsl} presentan diferencias en las tasas de subida y bajada, por lo que la relación entre las velocidades de entrada y salida de los \textit{router} también depende de la dirección del flujo de información. Dicha relación de velocidades, también se presenta en la interconexión de grandes \gls{isp} o en \gls{ixp} con tasas mayores, incluso en redes móviles donde los recursos de este tipo son todavía más escasos cuando la cantidad de usuarios de una celda es muy grande.

Estas diferencias entre las velocidades de entrada y de salida producen cuellos de botella donde puede ocurrir la pérdida de paquetes. Los \textit{router} utilizan \textit{buffer} para reducir las pérdidas de paquetes absorbiéndolos cuando estos no pueden ser reenviados en ese preciso instante, también, se utilizan como instrumentos que ayudan a mantener los enlaces con un alto grado de utilización en casos de congestión.

\section{Dimensionado}

Desde 1994, en \cite{buffers6} se propuso la denominada \textit{rule of thumb} o también llamada \gls{bdp}, la cual fue aceptada por muchos investigadores durante varios años, con el fin de determinar el tamaño de los \textit{buffer} en los nodos de una red. Esta regla, se describe en la ecuación \ref{eq_rule_of_thumb}, la cual define el tamaño del \textit{buffer}, $ B $, como el producto del ancho de banda del enlace, $ C $, por el retardo de ida y vuelta, \gls{rtt}. La ecuación \ref{eq_rule_of_thumb} se obtuvo utilizando $ 8 $ flujos \gls{tcp} en un enlace de $ 40 \; Mbps $, que en la actualidad no son datos representativos del tráfico en una red. Por este motivo, hoy en día no resulta un método factible debido al aumento de la cantidad de memoria necesaria con anchos de banda más grandes, por ejemplo, con una capacidad de $ 40 \; Gbps $, y un \gls{rtt} de $ 250 \; ms $, se obtendría un tamaño del \textit{buffer} de $ 1.25 \; Gbytes $ que es un tamaño muy grande (ver Ejemplo \ref{eje:rule_of_thumb}), además, no se tuvo en cuenta el caso de flujos con \gls{rtt} diferentes.

\begin{equation}
	\label{eq_rule_of_thumb}
	B=C \times RTT
\end{equation}

En $ 2004 $, esta regla fue puesta en duda, por el llamado \textit{Stanford model} \cite{buffers7} o también denominado \textit{small buffer} \cite{buffers5}, que reduce el tamaño del \textit{buffer}, dividiéndolo por la raíz cuadrada del número de flujos \gls{tcp}, $ N $, como se muestra en la ecuación \ref{eq_stanford_model}. Esto se debe a que la ausencia de sincronización entre los flujos permite realizar una aproximación. Este nuevo modelo se realiza bajo el supuesto de que la duración de los flujos es larga y el número de flujos es lo suficientemente grande como para considerarlos asíncronos e independientes.  

\begin{equation}
	\label{eq_stanford_model}
	B=\frac{C \times RTT}{\sqrt{N}}
\end{equation}

\begin{ejemplo}[La \textit{rule of thumb} y el \textit{Stanford model}][eje:rule_of_thumb]

La cantidad de memoria necesaria para un \textit{buffer}, según la \gls{bdp}, para una capacidad de $ 40 \; Gbps $, y un \gls{rtt} de $ 250 \; ms $, es de:

\begin{equation}
     \nonumber B=40 \; Gbps \times 250 \; ms = 10 \; Gbits = 1.25 \; Gbytes
\end{equation}

De acuerdo con el \textit{Stanford model} y suponiendo $100$ flujos, la cantidad de memoria necesaria sería:

\begin{equation}
     \nonumber B=\frac{40 \; Gbps \times 250 \; ms}{\sqrt{100}} = 1 \; Gbits = 125 \; Mbytes
\end{equation}

\end{ejemplo}

Debido al modelo propuesto por \cite{buffers7} se generó una serie de investigaciones en este ámbito. En \cite{buffers10} se propuso la utilización de \textit{buffer} todavía más pequeños, denominados \textit{tiny buffer}, que consideran que un tamaño de entre $ 20 $ y $ 50 $ paquetes (que equivale a algunas decenas de $ Kbytes $) es suficiente como para alcanzar una utilización del enlace de entre el $ 80\% $ y el $ 90\% $. Esto, basado en el hecho que los flujos no están sincronizados y el tráfico no presenta ráfagas. Sin embargo, muchos de los flujos \gls{ip} en tiempo real comúnmente tienen un comportamiento de ráfagas como por ejemplo el \textit{streaming} de video, esto deja un elemento de incertidumbre en cuanto a los modelos de dimensionado de \textit{buffer} para el tráfico de las aplicaciones que utilizamos hoy en día.

Por otro lado, son pocos los trabajos que consideran servicios de tiempo real, probablemente por el hecho de que gran parte del tráfico de Internet es \gls{tcp}, pero hoy en día, los servicios interactivos y aplicaciones multimedia tienen una demanda cada vez más grande \cite{gtc5}. En \cite{buffers9} y \cite{buffers2} se ha considerado un tráfico combinado de \gls{tcp} y \gls{udp} utilizando \textit{buffer} pequeños, descubriendo una región anómala, en la que las pérdidas de paquetes de \gls{udp} crecen con el aumento del tamaño del \textit{buffer} mientras que el \textit{throughput} de \gls{tcp} se mantiene.

En \cite{buffers4} se presentó una simulación mediante \gls{ns}, basada en una topología en árbol con $ 18 $ nodos y enlaces con una capacidad de $ 50 \; Mbps $, que muestra las variaciones de la pérdida de paquetes en función del tamaño de los \textit{buffer} para diferentes políticas de tráfico con la finalidad de mejorar el \textit{Stanford model}.

En general, se dice que un \textit{buffer} es un espacio de memoria en el cual se puede almacenar un determinado número de paquetes. Entonces, la cantidad de paquetes que se puede almacenar en un \textit{buffer} depende del tamaño, tanto de la memoria del \textit{buffer} como del paquete. Por este motivo, algunos investigadores y fabricantes realizan el dimensionado de los \textit{buffer} en términos de $ bytes $, mientras que otros, lo hacen en paquetes como se puede observar en \cite{buffers1}.

\section{Disciplinas de gestión de colas}

A medida que un sistema se congestiona, el retardo del servicio en el sistema aumenta, en estos casos, la probabilidad de tener un deterioro en la calidad, puede llegar a ser inaceptable para ciertos servicios con estrictos requerimientos de \gls{qos}. Por esta razón, la relación entre la congestión y el retardo es esencial para el diseño de algoritmos de control de congestión eficaces \cite{buffers20}. Las disciplinas de gestión de colas son herramientas que administran flujos de datos mediante determinadas políticas. Los sistemas operativos actuales, tanto de \textit{host} como de \textit{router}, implementan diversas técnicas para la gestión de los \textit{buffer} de las interfaces de red, estas técnicas pueden ser clasificadas como disciplinas de colas basadas en clases \gls{cbq} o colas de prioridad \gls{pq}.

Los \textit{buffer} de tipo \textit{drop-tail} son un ejemplo muy utilizado de colas \gls{pq}. Estos elementos encolan un paquete o $ byte $ si la cantidad de paquetes o $ bytes $ es menor que el tamaño máximo del \textit{buffer}, de lo contrario el paquete o $ byte $ es descartado. Un ejemplo de este tipo son los \textit{buffer} \gls{fifo}, en los cuales, el orden en que se almacena la información está asociado al orden de llegada de los datos, es decir, el primer dato en llegar, será el primero en ser transmitido en el momento que el \textit{router} tenga la capacidad de hacerlo. El tamaño de este tipo de cola puede ser definido en número de paquetes o en $ bytes $. 

Un \textit{buffer} \gls{pq} puede tener diversas colas donde los paquetes se van almacenando, cada cola tiene una prioridad diferente en función de determinadas políticas, las colas de menor prioridad podrán enviar paquetes solo si, las colas de mayor prioridad están vacías. Un ejemplo de este tipo de \textit{buffer} es \gls{fifo}-fast (ver Figura \ref{fig:fifo_fast}), el cual es un tipo de cola que puede ser configurada en cualquier interfaz de red en los sistemas operativos Linux \cite{linux1}. Este \textit{buffer} se compone de tres colas \gls{fifo} con distintas prioridades, donde la máxima prioridad la tiene la cola $ 0 $ y la mínima prioridad la cola $ 2 $. Los paquetes encolados en la $ 0 $ serán los primeros en ser procesados, los de la $ 1 $ serán procesados cuando no haya paquetes en la $ 0 $ y los paquetes asignados a la $ 2 $ serán procesados cuando no haya paquetes en las colas $ 0 $ ni $ 1 $. La asignación de un determinado paquete a una cola específica se realiza por medio del campo \gls{tos} de la cabecera \gls{ip}.

\InsertFig{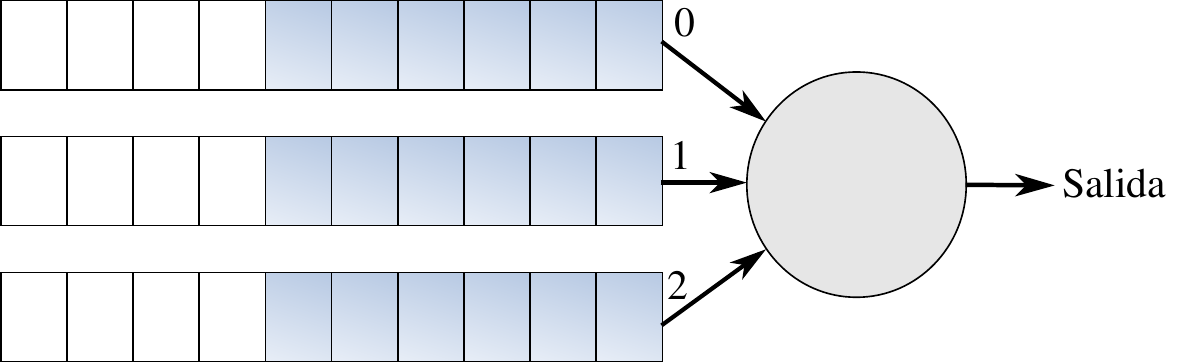}{fig:fifo_fast}{Descripción de un \textit{buffer} FIFO-fast.}{}{0.9}{}{}

Las colas de tipo \textit{drop-tail} tienden a penalizar los tráficos a ráfagas y a causar sincronización global en flujos \gls{tcp} debido a que los nodos reducirán, al mismo tiempo, la tasa de transmisión cuando se presenta la pérdida de paquetes. A pesar de esto, los \textit{drop-tail} \gls{fifo} abarcan una gran parte de las colas utilizadas en Internet debido a que son muy fáciles de implementar \cite{buffers15}, sin embargo, agravan las limitaciones de los esquemas de control de congestión de los terminales, como sucede en \gls{tcp}.

También, existen disciplinas de colas activas, \gls{aqm}, que suelen evitar este tipo de problemas, ya que descartan o marcan los paquetes para ser descartados probabilísticamente, antes de que la cola esté llena. La primera propuesta completa de \gls{aqm} fue \gls{red} \cite{buffers13}, el cual fue desarrollado para \gls{tcp} mediante el reemplazo de la colas \textit{drop-tail}. Los principales objetivos de \gls{red} fueron detectar la congestión cuando esta se está iniciando, lograr una equidad entre los flujos a ráfagas que tienen comportamientos diferentes, controlar la latencia, la sincronización global \cite{buffers16}, reducir al mínimo la pérdida de paquetes y proporcionar altos niveles de utilización del enlace.

Existe una gran cantidad de implementaciones diferentes de \gls{red} \cite{buffers21}, pero en general, se puede decir que se comporta como un \textit{buffer} \gls{fifo} cuando la cantidad de paquetes es menor a cierto umbral, por lo tanto, si el \textit{buffer} está casi vacío o por debajo de dicho umbral, se aceptan todos los paquetes entrantes. Cuando el tamaño de la cola crece por encima del umbral, la probabilidad de que un paquete sea descartado también crece. Cuando la ocupación del \textit{buffer} supera el umbral, los paquetes son descartados o marcados probabilísticamente. Cuando el \textit{buffer} está lleno, la probabilidad ha alcanzado el valor de $ 1 $ y todos los paquetes entrantes se eliminan, ver Figura \ref{fig:red}.

\InsertFig{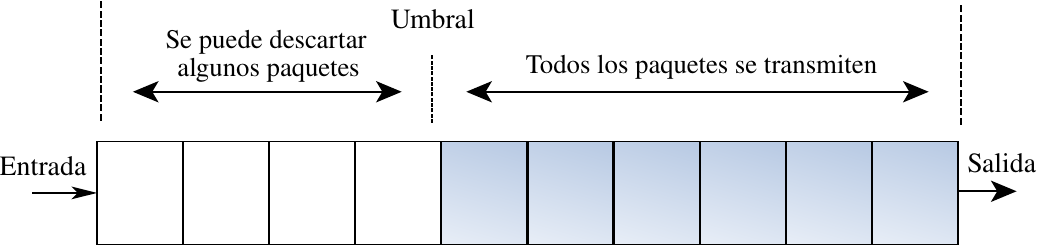}{fig:red}{Descripción de un \textit{buffer} tipo RED.}{}{}{}{}

El principal problema de esta técnica es la dificultad del ajuste óptimo de sus parámetros para un adecuado funcionamiento \cite{buffers17, buffers18} ya que es muy sensible a las condiciones de la red. Otro problema es que utiliza la longitud de la cola como una medida de su rendimiento y como un indicador de congestión, produciendo un deterioro en el \textit{throughput} y el retardo con el aumento del tráfico \cite{buffers19}, debido a que se obtendrá una alta tasa de pérdidas y un retardo grande cuando hay congestión.

Por otra parte, cuando los \textit{buffer} de los nodos de la red se encuentran llenos, las redes de conmutación de paquetes pueden causar valores muy elevados de latencia y de \textit{jitter}, deteriorando el rendimiento global de la red. A este fenómeno se le conoce como \textit{bufferbloat} (como se ha mencionado con anterioridad) y su efecto es más pronunciado cuando los \textit{buffer} son más grandes. Sin embargo, existen diversas técnicas de \gls{aqm} o variantes de \gls{red}, las cuales son capaces de mantener la longitud de la cola más pequeña, incluso algoritmos de planificación de la \gls{qos}, que previenen el \textit{bufferbloat} y reducen la latencia. Sin embargo, estas implementaciones de \textit{buffer} requieren mayor procesamiento y consumo de recursos para identificar tipos de tráfico, realizar mediciones de parámetros como \gls{rtt} o contar flujos, además, más memoria para gestionar diferentes colas. Por esto, en los \textit{router} de acceso no es común que se implementen este tipo de técnicas.

Se podría seguir citando una gran cantidad de estudios relacionados a técnicas de gestión de tráfico o algoritmos de planificación de colas, sin embargo, este tipo de técnicas ayudan a mantener cierto nivel de \gls{qos} cuando la utilización del enlace es alta y usualmente se implementan en \textit{router} de gama alta debido al consumo de recursos, los cuales se usan en el núcleo de la red. Como se mencionó en la sección \ref{cha:Introduccion}, la idea fundamental del libro es presentar los problemas de congestión cuando la utilización de los enlaces es media, valorando la influencia de los \textit{buffer} \gls{fifo} (sencillos de implementar y presentes en la mayoría de equipos comerciales de acceso de gama media y baja) en la \gls{qos} para casos de uso a nivel residencial, pequeñas empresas o grupos de usuarios.

\section{El rol del \textit{buffer} en la QoS}

En la actualidad existe un gran número de usuarios y dispositivos utilizando servicios multimedia que generan una cantidad de tráfico significativa en Internet \cite{games3, camera1} y la expectativa de crecimiento en el uso de aplicaciones multimedia indica que esta tendencia se incrementaría en los próximos años. Los servicios multimedia como la videoconferencia, la videovigilancia, la \gls{p2ptv} o los juegos \textit{online} alcanzan niveles de tráfico considerables para los \gls{isp} \cite{games2}. Esta carga de tráfico se puede considerar relativamente alta y se combina con el hecho de que este tipo de tráfico presenta características de comportamiento diferentes en comparación con otros servicios como web, \textit{email} incluso \gls{ftp} (o servicios en la nube con protocolos similares para la transferencia y sincronización de archivos).

Al mismo tiempo el tráfico generado por cada servicio depende de la naturaleza de la información que se transporta y de su tamaño. Como se describirá en el capítulo \ref{cha:Servicios Multimedia}, algunas de las aplicaciones multimedia generan tráfico a ráfagas cuando mucha información tiene que ser transmitida en un tiempo muy corto. Estas ráfagas pueden congestionar los dispositivos de red si la cantidad de paquetes es significativa con respecto al tamaño del \textit{buffer} de los dispositivos. Por otro lado, algunas aplicaciones trabajan para generar tráfico alisado \cite{smooth2, ls11}, con el objetivo de proveer un cierto nivel de \gls{qos} y una mejor experiencia al usuario sin ser perjudicial para la red, pero con el coste de un incremento en la capacidad de procesamiento \cite{ls12}.

El tamaño de los paquetes generados por estas aplicaciones puede variar entre los diferentes servicios de Internet, mientras algunos generan paquetes de tamaños pequeños que alcanzan unas pocas decenas de $ bytes $ (por ejemplo \gls{voip}) otros usan paquetes de mayor tamaño (por ejemplo videoconferencia). Sin embargo, el tamaño del \textit{buffer} y el ancho de banda disponible para soportar dichos servicios se mantienen en los mismos valores, por esto algunos enlaces de acceso podrían presentar problemas de congestión.

\subsection{El desbordamiento del \textit{buffer}}

Los \textit{buffer} pueden entrar en congestión por dos motivos principales: cuando la tasa de entrada es mayor a la tasa de salida, es decir el ancho de banda se agota y la utilización del enlace es alta, o bien, por problemas de dimensionado en la implementación de los dispositivos de red. Como se puede observar en la Figura \ref{fig:buffer}, el tiempo que se requiere para congestionar un \textit{buffer} está relacionado a la tasa de llenado, la cual está dada por la relación de las tasas de entrada y salida \cite{yo1, ls13}. 

\InsertFig{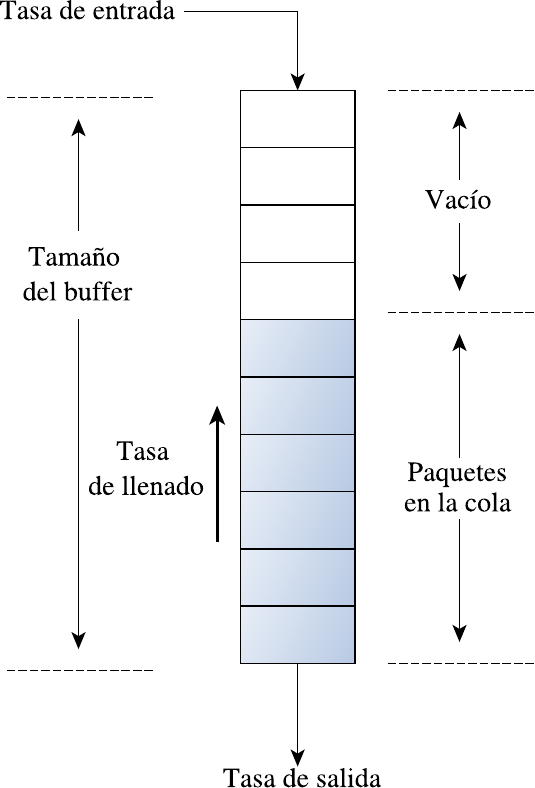}{fig:buffer}{Principales características de los \textit{buffer}.}{}{}{}{}

Se puede definir $ R_{in} $ y $ R_{out} $ como las tasas de entrada y salida respectivamente, además, se define $ R_{fill} $ como la tasa en la cual el \textit{buffer} se llena cuando $ R_{in} $ es más grande que $ R_{out} $ ($ R_{fill}=R_{in}-R_{out} $). Entonces, cuando un tráfico a ráfagas es generado en la red, la tasa de llenado del \textit{buffer} es muy alta, y en esos momentos, se puede dar una pérdida de paquetes, esto puede producirse incluso cuando la utilización del enlace es media o baja. Este fenómeno se puede presentar porque la longitud de la ráfaga es cercana al tamaño del \textit{buffer}, ya que este puede entrar en congestionamiento más fácilmente, también puede darse, cuando la longitud de la ráfaga es mayor que el tamaño del \textit{buffer}, en cuyo caso, la pérdida de paquetes será mucho más probable, ver Ejemplo \ref{eje:desbordamiento-rafagas}. Es cierto que muchas aplicaciones implementan mecanismos para generar un tráfico suavizado, pero el tráfico global de Internet tiene un comportamiento a ráfagas en todas las escalas \cite{bursty1}.

\begin{ejemplo}[Desbordamiento del \textit{buffer} debido a ráfagas][eje:desbordamiento-rafagas]

Si una ráfaga con $20$ paquetes de video llega a un \textit{buffer} que tiene un tamaño de $15$ paquetes, hay $5$ paquetes que se descartarán ya que no hay suficiente espacio en el \textit{buffer} para almacenarlos a todos.

\end{ejemplo}

En algunos escenarios, cuando se presentan problemas de congestión de la red, una práctica común puede ser aumentar el ancho de banda en la red local. Por esta razón, muchas compañías cambian sus dispositivos de red interna (por ejemplo, cambiando de una velocidad menor a una mayor) tratando de resolver los problemas de congestión. Pero, si $ R_{out} $ se mantiene en el mismo valor y $ R_{in} $ se cambia a una tasa mayor, la tasa de llenado del \textit{buffer} será mayor ($ R_{fill} $) en la nueva red, como se muestra en el Ejemplo \ref{eje:desbordamiento-tasa}. Por esta razón, en casos de tráficos a ráfagas, el \textit{buffer} se congestionará más rápidamente. Así, en ciertos casos, este aumento de la velocidad en la red interna puede producir una respuesta peor de la red, de tal manera, que esta mejora se convierte en un fracaso. En definitiva, la relación entre las velocidades de la red local y el acceso a Internet, y la relación entre el tamaño del \textit{buffer} y la longitud de la ráfaga son, de hecho, parámetros importantes que no pueden ser descuidados.

\begin{ejemplo}[Desbordamiento del \textit{buffer} debido a la tasa de entrada][eje:desbordamiento-tasa]

Una red envía datos a una tasa constante de $5 \; Mbps$ por un enlace de acceso a Internet de $5 \; Mbps$. En este caso, la tasa de llenado es de:

\begin{equation}
    \nonumber R_{fill} = 5 \; Mbps - 5 \; Mbps = 0 \; Mbps
\end{equation}

Esto significa que el \textit{buffer} no se está llenando y puede gestionar la entrega del tráfico sin ningún problema. Si la red incrementa la tasa de envío a $10 \; Mbps$ y el enlace de acceso se mantiene en el mismo valor, la nueva tasa de llenado sería:

\begin{equation}
    \nonumber R_{fill} = 10 \; Mbps - 5 \; Mbps = 5 \; Mbps
\end{equation}

Inicialmente el \textit{buffer} encolará los paquetes que no se pueden transmitir en ese momento a una tasa de $5 \; Mbps$ hasta que el espacio de memoria lo permita. Si la tasa de entrada se mantiene, el \textit{buffer} se llenará completamente y descartará el $ 50\% $ del tráfico.

\end{ejemplo}

Por otro lado, la generación de tráfico a ráfagas por parte de las aplicaciones no es el único motivo por el cual un \textit{buffer} puede producir una pérdida de paquetes. Usualmente, el tráfico de las aplicaciones comparte un enlace con otros tipos de tráfico de servicios con comportamientos diferentes en cuanto a la generación de sus paquetes. Dichos flujos de datos pueden ser generados por el mismo \textit{host}, o bien, por la convergencia de flujos de diversos equipos hacia un enlace en común. Esta combinación de los flujos de tráfico que comparten un mismo enlace, puede producir ráfagas que repercutan más drásticamente en la tasa de llenado de los \textit{buffer} de la red. Además, dicha ráfaga puede llegar a contener una cantidad de paquetes que supere el tamaño de ciertos \textit{buffer}, lo cual repercutiría negativamente en la \gls{qos} de ciertas aplicaciones más susceptibles.

\subsection{Influencia del \textit{buffer} en diferentes servicios}

Existen muchas publicaciones científicas relacionadas con la influencia del \textit{buffer} en diferentes servicios y aplicaciones que muestran cómo la \gls{qos} es afectada por el comportamiento del \textit{buffer}, el cual está principalmente definido por su tamaño y sus políticas de gestión. En estos casos, el conocer las características técnicas y funcionales de estos dispositivos se convierte en un aspecto fundamental. Este conocimiento puede ser útil para diversas aplicaciones y servicios con la finalidad de decidir y gestionar la forma en que el tráfico es generado. Además, se pueden aplicar ciertas técnicas de gestión de paquetes como por ejemplo, multiplexar un cierto número de paquetes pequeños dentro de uno más grande, o por el contrario, aplicar la fragmentación o incluso suavizar el tráfico, de acuerdo a cada escenario \cite{ls14, ls15}.

La manera en que se estudia la influencia del \textit{buffer}, para el tráfico multimedia, es determinando las características de \gls{qos}, basado en parámetros bien conocidos de la red, por ejemplo, el retardo, el \textit{jitter} y la pérdida de paquetes. También se usan evaluaciones de la calidad subjetiva para determinar la percepción de los usuarios para ciertos servicios. Como se mencionó en el capítulo \ref{cha:Calidad de Servicio}, el \textit{E-Model} de la \gls{itu} \cite{emodel, mos1}, presenta un procedimiento con el objetivo de calcular el \gls{mos}, el cual es útil en el planeamiento de transmisión de red. Otros autores \cite{games5}, han desarrollado un modelo similar para juegos \textit{online} con base en el retardo y el \textit{jitter} y en general se puede afirmar que existe un modelo para cada tipo de comunicación. A continuación se mencionan algunos de dichos estudios.

\begin{itemize}
    \item La influencia del \textit{buffer} en \gls{voip} se ha estudiado en \cite{gtc17}, donde se probaron tres políticas de \textit{buffer} diferentes (\textit{buffer} dedicado, grande y limitado en tiempo) con dos técnicas de multiplexión. Donde, cada política del \textit{buffer} del \textit{router} causó un comportamiento diferente en la pérdida de paquetes y también modificó la calidad de la voz, la cual se midió por medio del \textit{R-Factor} \cite{mos1}. En el mismo artículo, se estudió un método de multiplexión para flujos de \gls{voip}, en el cual se obtuvo una reducción del ancho de banda con el aumento del tamaño de los paquetes, lo que influye en la pérdida de paquetes dependiendo de la implementación del \textit{buffer} y su tamaño. En este caso, el tráfico nativo de \gls{voip} mostró un buen comportamiento cuando se usaron \textit{buffer} pequeños y medidos en $ bytes $, ya que en estos casos, los paquetes pequeños tienen menos probabilidad de ser descartados que los grandes.
    \item En \cite{gtc14}, los autores presentaron un estudio de simulación, de la influencia de un método de multiplexión en los parámetros que definen la calidad subjetiva de los juegos \textit{online} (principalmente retardo, \textit{jitter} y pérdida de paquetes). Los resultados muestran que los \textit{buffer} pequeños, presentan mejores características para mantener el retardo y el \textit{jitter} en valores adecuados, pero a costa de incrementar la pérdida de paquetes. Además, los \textit{buffer} cuyos tamaños se miden en paquetes también incrementan la pérdida de paquetes.
    \item Muchos dispositivos de las redes de acceso están diseñados para la transferencia de datos a granel \cite{gtc18}, como los servicios de correo, web o \gls{ftp}. Sin embargo, otras aplicaciones (por ejemplo, \textit{streaming} de video \gls{p2p}, juegos \textit{online}, etc.) generan altas tasas de paquetes pequeños, en estos casos, los \textit{router} podrían experimentar problemas para gestionar todos los paquetes. Por lo tanto, la capacidad de procesamiento se puede convertir en un cuello de botella en dichos dispositivos, si estos no pueden gestionar suficientes paquetes por segundo \cite{games4}. En este escenario, lo que sucede es que la tasa de salida disminuye.
    \item La generación de altas tasas de paquetes pequeños también ha sido observada en aplicaciones \gls{p2ptv} \cite{p2p10}, dichas aplicaciones además generan tráfico de video. En los casos en los que un tráfico mixto de paquetes pequeños y grandes atraviesa un \textit{buffer} medido en paquetes, los paquetes de video pueden verse penalizados por los paquetes pequeños ya que ambos tendrán la misma probabilidad de ser descartados, y como consecuencia, el comportamiento del \textit{peer} no será la esperada en una estructura \gls{p2p}.
\end{itemize}

\chapter{Los Servicios Multimedia}
\label{cha:Servicios Multimedia}
% \minitoc
En la actualidad los usuarios de Internet demandan una gran variedad de servicios multimedia, muchos de ellos con estrictos requerimientos de tiempo real. Además, la pandemia del virus COVID-$19$ (que inició en el año $2020$) ha forzado a millones de personas en todo el mundo a trabajar de manera remota, usualmente desde casa, haciendo uso extensivo de aplicaciones de videoconferencia sobre redes de acceso con mediana o baja capacidad, en muchos casos. También, destacan otras clases de aplicaciones, por ejemplo, la telefonía por Internet, que debido a la reducción de costes ha alcanzado un gran auge. Por otro lado, se pueden resaltar los sistemas de televigilancia con un alto grado de aceptación a nivel gubernamental, empresarial e incluso en el ámbito residencial. También, los sistemas de televisión y servicios interactivos forman parte de este tipo de aplicaciones y servicios. En este capítulo se describen algunas de las principales aplicaciones y el comportamiento de estas, en cuanto al tráfico generado en la red. En este sentido, se desea conocer el tamaño de los paquetes, si el tráfico es a ráfagas y el tamaño de dichas ráfagas, el tiempo entre paquetes, protocolos de transporte o incluso protocolos a nivel de aplicación que den una idea de cómo una determinada aplicación genera tráfico a la red. Esta información es útil para estimar el impacto que un determinado tipo de tráfico puede tener en la red. 

\section{VoIP}

El desarrollo de tecnologías de \gls{voip} ha tenido una gran aceptación por parte de empresas que buscan una reducción de costes para sus comunicaciones de voz; principalmente \gls{pymes} \cite{gtc8, gtc1}. \gls{voip} permite la transmisión de voz por medio de una red \gls{ip}, basándose en la digitalización de las señales de voz por medio de un \textit{codec}. Además, \gls{voip} hace uso de diversos tipos de técnicas para la señalización de la llamada, no habiendo un protocolo único definido en este ámbito. Uno de los protocolos más utilizados para este fin es \gls{sip} \cite{rfc3261}, también, se encuentran implementaciones normalizadas con H.323 o propietarias que se hacen públicas como \gls{iax}, y por último otras que no se hacen públicas como las de Skype.

Centrándose en \gls{sip}, este es uno de los protocolos con mayor impacto en la implementación de \gls{toip} \cite{gtc9, gtc1}. Dicho protocolo, se encarga de la señalización extremo a extremo de la comunicación y realiza los procedimientos necesarios para el establecimiento de la llamada, la modificación y la canalización de la comunicación \cite{gtc3}. Por otro lado, para la transmisión de datos en tiempo real, generalmente se hace uso del protocolo \gls{rtp}; dicho protocolo se encarga del control de la transmisión en las sesiones de aplicaciones multimedia y utiliza como protocolo de transporte \gls{udp}. La Figura \ref{fig:paquete_voip} muestra la estructura de paquetización de \gls{voip} entre dos equipos terminales y el Ejemplo \ref{eje:voip} describe cómo calcular el ancho de banda a nivel \gls{ip} de una transmisión de voz. El tráfico de este tipo de servicio tiene una distribución uniforme (no se presentan ráfagas de paquetes), cada equipo terminal realiza el envío de paquetes cada cierto tiempo, el cual está determinado por la paquetización y el \textit{codec} utilizado. 

\InsertFig{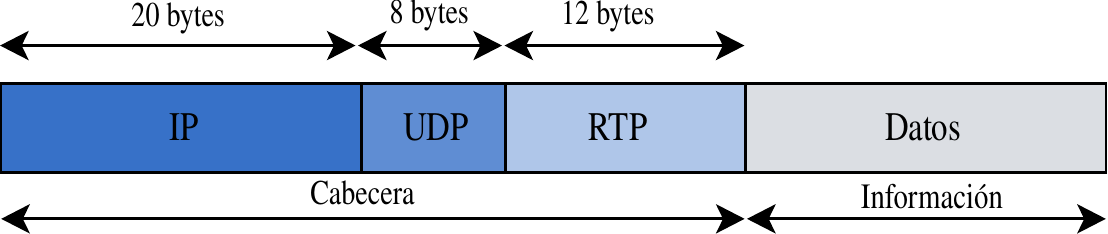}{fig:paquete_voip}{Descripción de un paquete VoIP transmitido entre dos estaciones de trabajo.}{}{0.9}{}{}

\begin{ejemplo}[Ancho de banda para un flujo \gls{voip}][eje:voip]

Al capturar una traza de tráfico \gls{voip}, en la cual los equipos terminales se configuraron con el \textit{codec} $ G$.$729 $ y con una cantidad de $ 2 $ muestras de voz por paquete, se observó que el tiempo medio de envío de paquetes es de $ 20 \, ms $ con una desviación estándar de $ \pm 0.62 \, ms $. Cada conexión de este flujo tienen un consumo de ancho de banda a nivel \gls{ip} de: 

\begin{equation}
    \nonumber BW=\frac{IP+UDP+RTP+Datos}{muestras \times 10ms}=\frac{60\times8}{20\times10^{-3}}=24 \; Kbps
\end{equation}

\end{ejemplo}

En este tipo de servicio la pérdida de paquetes y el retardo son parámetros importantes que determinan la \gls{qos}. En términos generales se dice que se ha perdido una trama \gls{voip} cuando esta no llega a tiempo para ser reproducida. Por este motivo, también tiene una gran influencia el \textit{jitter} que tenga la red.

Por otro lado, hay algunos estudios que relacionan el comportamiento de los usuarios (por ejemplo, a la hora de iniciar, cerrar o reiniciar una sesión en determinadas aplicaciones) y la estabilidad de la red \cite{skype8}, mostrando que los flujos \gls{voip} no solo consumen menos ancho de banda que los flujos \gls{tcp}, sino que también, son muy sensibles a la congestión cuando la red está altamente cargada. De esta manera, se sugiere que el comportamiento del usuario y el diseño de la aplicación van a jugar un papel cada vez más importante en el análisis de la infraestructura de red, en la distribución de los recursos de la red y el control de congestión. 

\section{Videovigilancia}

Las cámaras \gls{ip} han tenido un impacto importante en los mecanismos de seguridad a nivel empresarial y residencial, este tipo de equipos permite emitir video (utilizando técnicas de compresión de imagen) a través de Internet utilizando \gls{tcp}/\gls{ip}. Dentro de las funciones más usuales, se encuentran, por ejemplo, la activación mediante movimiento o sensores, control remoto y gestión a través de \gls{http}. Cada una de dichas funciones produce un efecto diferente en el tráfico generado a la red. El formato de imagen más usual es \gls{jpeg} con soporte para diferentes niveles de compresión. De manera muy general y para obtener alta calidad, se puede decir que una cámara de este tipo captura imágenes, las convierte a un formato \gls{jpeg} y las transmite a razón de $ 25 $/$ 30 $ \textit{frames} por segundo para \gls{pal}/\gls{ntsc}. Puede trabajar sin problemas sobre una red con ancho de banda de $ 10 \; Mbps $ o $ 100 \; Mbps $ \cite{axis2120}.

El comportamiento del flujo de datos difiere en función de la configuración que permita el fabricante para este tipo de dispositivos. Uno de los factores con mayor relevancia es el nivel de compresión que se defina para la imagen, ya que este define su tamaño (en $ bytes $) y afectará de forma directa a las características del flujo de paquetes en la red, y también influye en la calidad percibida por el usuario. 

El modelado de tráfico es muy útil para estimar y comprender un determinado flujo de datos. Una manera de modelar el tráfico de una cámara consiste en considerar que una cámara transmite en cada instante imágenes que tienen un tamaño diferente. Estas imágenes son enviadas a la red mediante un determinado número de paquetes en forma de ráfaga, el último paquete de cada ráfaga, tendrá un tamaño menor que los demás, mientras que el resto serán de $ 1500 \; bytes $, esto se aprecia en la Tabla \ref{table:cantidad_paquetes_camara}. Por lo tanto, los parámetros básicos del modelo serán el número de imágenes por segundo y el número de paquetes por imagen.

\begin{table}[htbp!]
	\centering
	\vspace{0mm}
	\scalebox{0.85}[0.9]{
	\begin{tabular}{ccc}
		\toprule
		\textbf{Resolución} & \textbf{Nivel de compresión} & \textbf{Cantidad de paquetes} \\
		\midrule 
		\rowcolor{row1} & $ 50 \; Kbytes $ & $ 25 $ \\ 
		%\cline{2-3}
		\rowcolor{row1} \multirow{2}{*}[6mm]{$ 704\times576 \; pixeles $} & $ 16 \; Kbytes $ & $ 10 $ \\ 
		\midrule[0pt]
		\rowcolor{row2} & $ 13 \; Kbytes $ & $ 9 $ \\ 
		%\cline{2-3}
		\rowcolor{row2} \multirow{2}{*}[6mm]{$ 352\times288 \; pixeles $} & $ 4 \; Kbytes $ & $ 3 $ \\ 
		\bottomrule
	\end{tabular} }
	\caption{Cantidad de paquetes por ráfaga en función de la compresión para una cámara IP AXIS 2120.}
	\label{table:cantidad_paquetes_camara}
\end{table}

Otra función que en la actualidad la mayoría de estas cámaras tienen, son sensores para determinar el movimiento y esto tiene un efecto en el tráfico de la red. La cantidad de paquetes que una cámara de este tipo envía a la red también depende del movimiento percibido por la cámara, en estos casos se observa que cuando hay mayor movimiento en la imagen percibida por la cámara, la cantidad de paquetes enviados lógicamente aumenta.

\section{Videoconferencia}

Los sistemas de videoconferencia se han extendido ampliamente gracias a las mejoras en las técnicas de compresión de imágenes y al aumento del ancho de banda de las las redes de acceso. En la actualidad, existen múltiples aplicaciones que permiten este tipo de servicio incluso en dispositivos con recursos limitados (por ejemplo, dispositivos móviles). 

La arquitectura de este tipo de servicios puede dividirse en modelos centralizados (cliente-servidor) o \gls{p2p}. Las arquitecturas centralizadas son una solución interesante para dispositivos portátiles como teléfonos inteligentes o tabletas, con una limitada capacidad de procesamiento y energía, ya que permite la reducción de estos aspectos en los clientes, mientras concentra el procesamiento en un nodo central con una alta capacidad. Algunos ejemplos de este tipo de sistemas son: Vidyo y Google plus hangout. En un sistema \gls{p2p} cada nodo actúa simultáneamente como cliente y servidor, permitiendo el intercambio directo de información entre los \textit{peer} interconectados. Este tipo de redes, aprovechan el ancho de banda de los usuarios por medio de la conectividad entre ellos mismos, y obtienen mejor rendimiento que con algunos métodos centralizados convencionales, cuando una cantidad de servidores es relativamente pequeña. Además es una técnica interesante para afrontar los problemas de escalabilidad de las redes centralizadas.

Al igual que en la videovigilancia, dentro de los aspectos fundamentales a la hora de modelar el tráfico del servicio se encuentra la codificación y las tecnologías de compresión utilizadas. La codificación tradicional se basa en que existen diferentes \textit{codec} que consiguen mayor calidad pero enviando más información. Algunos sistemas de videoconferencia utilizan tecnologías de compresión \gls{svc}, que son codificadores de video escalables y diseñados para incluir una mayor flexibilidad a los sistemas multimedia. La gran diferencia con los \textit{codec} tradicionales es que, incluyen \gls{avl}, generando un tráfico de salida en múltiples capas donde cada capa aumenta la calidad del video recibido por el usuario. Este enfoque escalable es adecuado para usuarios con ciertas restricciones de ancho de banda o con accesos con problemas de congestión, ya que en este tipo de entornos, se podría recibir una cantidad de capas menor, manteniendo el servicio a pesar de tener una calidad relativamente inferior \cite{video1}. También, un usuario con mejores prestaciones puede recibir más capas, mejorando la calidad en función de los recursos que dispone. Cada capa emplea predicción con compensación de movimiento e intra-predicción \cite{video2}. La ventaja de este sistema frente a los tradicionales, es que puede adaptarse a las condiciones de la red mediante el filtrado de capas en lugar de tener que cambiar de \textit{codec}. Las capas pueden ir en paquetes diferenciados, por lo que el filtrado solamente consiste en filtrar determinado tipo de paquetes.

En cualquier caso, la cantidad de paquetes que este tipo de aplicaciones envían a la red depende del tamaño del \textit{frame}, y este a su vez, depende del modelo de codificación utilizado y el movimiento del video. En \cite{video3}, se puede observar una comparación de la variabilidad del tráfico de diversas secuencias de \textit{Silence of the Lambs} y \textit{Star Wars IV} para tres tipos de suavizado de tráfico. Por este motivo, resulta a menudo inviable establecer un modelo de tráfico, siendo preferible la utilización de trazas de tráfico real. 

A continuación se mencionan algunos ejemplos de estudios en relación al análisis del tráfico de aplicaciones de videoconferencia.

\begin{itemize}
    \item Uno de los principales ejemplos de la videoconferencia (que por cierto, sigue una arquitectura \gls{p2p}) es Skype. En \cite{skype1}, se analizan las principales funciones de Skype como \textit{login}, \gls{nat}, el establecimiento de la llamada y el \textit{codec}. Algunos estudios se centran en caracterizar ciertas capas de la arquitectura, el comportamiento del protocolo \gls{p2p} y el tráfico de voz. En \cite{skype2}, los autores comentan que esta aplicación reacciona diferente ante la pérdida de trayectoria y la congestión de la red, además, que inunda la red con paquetes pequeños de señalización con la finalidad de mantener el servicio de forma eficaz, sin embargo, esta práctica puede resultar costosa desde el punto de vista de algunos dispositivos de red.
    \item Otros estudios \cite{skype3, skype4}, analizan la calidad de las llamadas de voz y la \gls{qoe}, proponiendo un modelo para cuantificar el nivel de satisfacción de los usuarios. Algunos \cite{skype5} han propuesto mecanismos para el control de la congetión en el tráfico de \gls{voip} de Skype.
    \item Con respecto a la capacidad de respuesta de las llamadas de video de Skype, los autores de \cite{skype6} midieron las variaciones del ancho de banda y llegaron a la conclusión de que el tiempo de respuesta es grande, cuando el ancho de banda se incrementa. Sin embargo, este estudio sólo tiene en cuenta el comportamiento transitorio de Skype, y no midió sistemáticamente su comportamiento estacionario cuando este es alcanzado. Además, los autores de \cite{skype7} caracterizaron los sistemas de control de velocidad y la calidad de las video llamadas, mostrando que dicha aplicación es robusta cuando las pérdidas de paquetes y los retardos de propagación son leves y puede utilizar de manera eficiente el ancho de banda de red disponible.
    \item Otro ejemplo de este tipo de sistemas es Vidyo, el cual es una alternativa propietaria cuya ventaja frente a otras soluciones de videoconferencia es la utilización de \gls{avl}. En \cite{gtc16}, se presenta un estudio que muestra una comparativa de la adaptación del tráfico a la red, para Skype y Vidyo, cuando cambian ciertos parámetros de la red. Las pruebas presentadas se realizaron en un entorno controlado de laboratorio en el cual se varía el ancho de banda, el retardo y la pérdida de paquetes. Los resultados muestran que Vidyo es capaz de detectar rápidamente variaciones en la red y puede adaptar su tráfico según corresponda. Ante los cambios en la red, Vidyo reacciona variando el ancho de banda generado, el tamaño de los paquetes y el tiempo entre paquetes.
\end{itemize}

\section{Video \textit{streaming}}

El crecimiento a nivel mundial en el acceso a Internet por parte de los usuarios móviles ha generado el desarrollo de diversas aplicaciones y nuevos modelos de negocios utilizando servicios de \textit{streaming}, entre ellos se encuentran muchas plataformas de redes sociales, la radio y la televisión por Internet (por mencionar algunos) \cite{gtc5} y \cite{otros3}. Este tipo de servicios basa su funcionamiento en la transmisión \textit{streaming}. El \textit{streaming} consiste en la distribución de audio o video por Internet, esta palabra hace referencia a una transmisión en forma continua, sin interrupciones y sin la necesidad de descargas previas. 

El comportamiento de este tipo de tráfico está relacionado con la configuración del proveedor del servicio, la selección de protocolos para el transporte, el control y la compresión. En este ámbito, es común para aplicaciones en tiempo real, el uso de \gls{udp} para el transporte y \gls{rtp} para el control de sesión en tiempo real, además, en cuanto a la compresión muchos servicios utilizan \gls{mpegts}. También, existen estudios que han caracterizado y medido el impacto en el tráfico de la red para aplicaciones de video \textit{streaming} \cite{p2p14}. 

El tráfico \textit{streaming} no tiene un comportamiento uniforme, presenta ráfagas que tienen grandes diferencias en cuanto a los tiempos entre los inicios entre ellas. Las ráfagas producidas durante la transmisión presentan variaciones en la duración de las mismas, sin embargo, si se utiliza \gls{ts} los paquetes tienen un tamaño máximo entorno a $ 1370 \, bytes $. Las ráfagas se caracterizan por mantener un período de inactividad antes del inicio de la próxima ráfaga. En general, el tamaño de los paquetes es el mismo en todos los casos, ya que cada \textit{stream} elemental tiene un tamaño fijo de $ 188 \, bytes $ \cite{h222} y los paquetes \gls{ip} deben contener múltiplos de estos, por lo tanto el mayor tamaño posible si consideramos un \gls{mtu} de $ 1500 \; bytes $ sería de $ 1370 \, bytes $. Sin embargo, al igual que en los sistemas de videovigilancia, resulta inviable establecer un modelo de tráfico, dadas las características del tráfico, siendo mejor utilizar trazas de tráfico real.

En cuanto a la percepción de los usuarios, mientras los servicios de video \textit{streaming} no tienen unos requerimientos temporales excesivos y pueden sufrir retardos sin un deterioro de la calidad percibida, el \textit{streaming} de televisión es un servicio más sensible a los retardos debido a sus características, como se ve por ejemplo en los programas en vivo.

\section{P2P-TV}

\gls{iptv} es un servicio interactivo en tiempo real que tiene un impacto importante en el tráfico de la red, ya que los requisitos de ancho de banda, derivados del envío de flujos \textit{unicast} a cada usuario, es bastante caro \cite{p2p14}. Los sistemas \gls{p2ptv} son una técnica de difusión de contenidos usando una arquitectura \gls{p2p}. Estos sistemas son una solución práctica a los problemas de escalabilidad de las redes \gls{iptv}, debido a que el consumo de recursos de ancho de banda es menor, y por lo tanto disminuye su coste \cite{p2p13}.

Sin embargo, los sistemas \gls{p2ptv} tienen un comportamiento particular que se debe tener en cuenta ya que generan una gran cantidad de paquetes pequeños. En \cite{p2p17}, se caracterizó el tráfico de una de las aplicaciones más populares en sistemas \gls{p2ptv}, en dicho estudio, los autores afirman que las aplicaciones de este tipo, generan su propio patrón de tráfico, pero tienen en común que este se compone de una combinación de paquetes pequeños y paquetes de gran tamaño, además, que los de menor tamaño corresponden a paquetes de señalización, mientras que los grandes a paquetes de video.

Una aplicación de este tipo de servicios es SopCast, la cual utiliza \gls{udp} como protocolo de transporte. Esta aplicación tiene un \textit{overhead} bastante alto, ya que, cerca del $ 60\% $ de los paquetes que generan son de señalización y solo el $ 40\% $ corresponde a datos de video \cite{p2p1}. El comportamiento del tráfico de un \textit{peer} se caracterizó en \cite{gtc15}, descubriendo que un cliente SopCast envía un paquete \gls{udp} de confirmación por cada paquete de video recibido, generando una cantidad considerable de paquetes pequeños en el enlace ascendente del cliente, además, el flujo de paquetes de confirmación influye negativamente en el tráfico de video que dicho \textit{peer} envía hacia otros, ya que ambos flujos competirán en el enlace de subida.

\chapter{El \textit{Buffer} y las ráfagas}
\label{cha:Buffer y rafagas}
% \minitoc
En este capítulo se presenta un análisis de las características de los \textit{buffer} (especialmente su tamaño y la pérdida de paquetes) en los dispositivos de acceso. En particular se estudia cómo estas características pueden afectar a la calidad de las aplicaciones multimedia cuando estas generan tráfico a ráfagas en la red local. 

En primer lugar, se muestra un escenario con flujos de tráfico a ráfagas en el cual se ha escogido un servicio de videovigilancia con cámaras \gls{ip} para una red de \gls{pymes}, en dicho escenario los flujos comparten el mismo enlace de acceso. Además, se presenta cómo el aumento de capacidad de la red interna podría causar el desbordamiento de los \textit{buffer} (cuando se mantiene la misma capacidad del enlace de salida) y producir una cantidad significativa de pérdida de paquetes que podrían deteriorar la \gls{qos}.

En segundo lugar, este capítulo muestra que la naturaleza a ráfagas de aplicaciones como la videovigilancia puede perjudicar su \gls{qos}, especialmente cuando cierto número de ráfagas se solapan, ya que las ráfagas de pérdidas de paquetes en un \textit{buffer} no se producen únicamente por aplicaciones que generan tráfico a ráfagas, sino, que también pueden ser causadas por el solapamiento de diferentes flujos en un enlace sensible. Para abordar este tema, se han planteado una serie de pruebas con el objetivo de caracterizar el problema que se puede presentar debido al incremento de la demanda de capacidad en este tipo de escenarios. En algunos casos, especialmente cuando se presentan aplicaciones que generan tráfico a ráfagas, el incremento de la capacidad de la red podría conducir a un deterioro de la calidad. Como conclusión, se muestra que en estos casos la principal causa de la degradación de la calidad, en caso de no sobrepasar la capacidad del enlace, se debe al desbordamiento del \textit{buffer}, y que depende de la relación entre el ancho de banda de la red interna y la de acceso.
 
\section{Escenario de red propuesto}

Como se ha mencionado con anterioridad, se ha seleccionado un servicio de videovigilancia que utiliza cámaras \gls{ip}, las cuales se pueden acceder mediante un navegador de Internet convencional. El escenario utilizado para las pruebas se muestra en la Figura \ref{fig:test_camera}, en la cual varias comunicaciones de videovigilancia comparten el mismo enlace de acceso a Internet hacia el centro de vigilancia. El principal objetivo de las pruebas es determinar la tasa de pérdida de paquetes en el tráfico a ráfagas combinado para diferentes tamaños de \textit{buffer}, y observar los diferentes resultados cuando la capacidad de los enlaces entre las cámaras y el dispositivo de acceso a Internet cambia de $ 10 \; Mbps $ a $ 100 \; Mbps $. 

\InsertFig{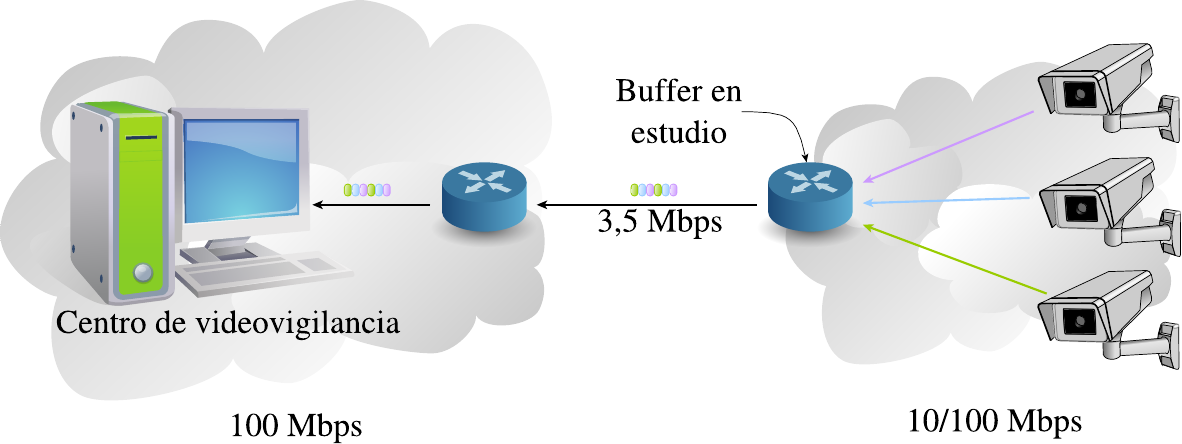}{fig:test_camera}{Escenario para las pruebas de uno, dos y tres flujos de datos de cámaras IP.}{}{0.9}{}{}

La prueba se repite utilizando tráfico de $ 1 $, $ 2 $ y $ 3 $ cámaras, donde cada una transmite a una tasa media de $ 1 \; Mbps $. La capacidad del enlace de acceso se limita a $ 3.5 \; Mbps $, dicho valor se ha seleccionado con el fin de establecer la utilización al $ 85 \% $ de la capacidad del enlace cuando las tres cámaras transmiten al mismo tiempo. Además, estableciendo la utilización del enlace en dicho valor, se puede asegurar que la pérdida de paquetes es causada por la relación entre las características del tráfico y el comportamiento del \textit{buffer} del \textit{router} y no por la escasez del ancho de banda.

\section{Tráfico utilizado}

Con la finalidad de desarrollar las pruebas descritas anteriormente, se han utilizado trazas reales de aplicaciones de videovigilancia, las cuales fueron capturadas en escenarios reales, para luego ser generadas en \gls{ns}, utilizando los mismos tamaños de paquetes y tiempos entre paquetes. La metodología utilizada para la captura de tráfico se ilustra en la Figura \ref{fig:test_traffic}, en la cual se incluye un \textit{sniffer} en la mejor ubicación para que no degrade el rendimiento de las aplicaciones \cite{capturing1}.

\InsertFig{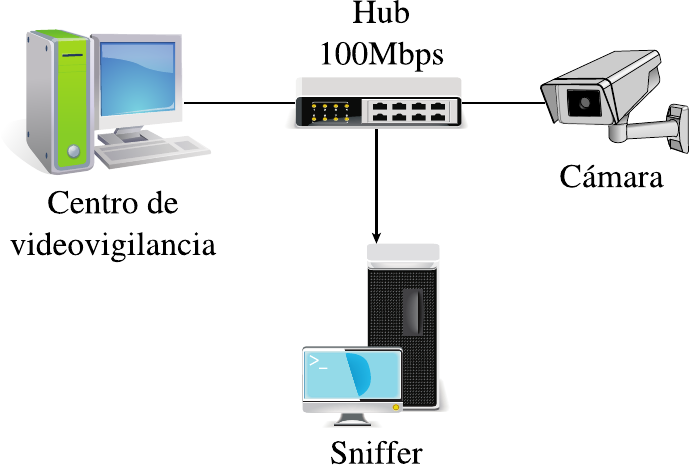}{fig:test_traffic}{Escenario para la captura del tráfico para un sistema de videovigilancia.}{}{}{}{}

Las trazas del tráfico de videovigilancia se obtuvieron utilizando una cámara \gls{ip} bien conocida en el mercado (AXIS $ 2120 $). Este tipo de tráfico es particularmente a ráfagas, en la Tabla \ref{table:camera_packets} se muestra la relación entre el nivel de compresión de video y la cantidad de paquetes por ráfaga para dos diferentes resoluciones cuando el ancho de banda de la cámara se establece en $ 1 \; Mbps $. Para todas las pruebas realizadas se han seleccionado trazas con una resolución de $ 704 \times 576 \; px $ y una compresión de $ 32 \; Kbytes $. El tiempo medio entre las ráfagas es de $ 0.278 \; s \; \pm \; 0.06 \; s $, la cantidad de paquetes por ráfaga es de $ 26 $ y el tamaño de los paquetes es de $ 1500 \; bytes $.

\begin{table}[htbp!]
	\centering
	\vspace{0mm}
	\scalebox{0.85}[0.9]{
	\begin{tabular}{ccc}
		\toprule
		\textbf{Resolución} & \textbf{Nivel de compresión} & \textbf{Cantidad de paquetes} \\
		\midrule
		\rowcolor{row1} & $ 50 \; Kbytes $ & $ 41 $ \\ 
%		\rowcolor[gray]{0.9}$ 704\times576 \; pixeles $ & $ 32 \; Kbytes $ & $ 26 $ \\
		\rowcolor{row1}$ 704\times576 \; pixeles $ & $ 32 \; Kbytes $ & $ 26 $ \\
		\rowcolor{row1} & $ 16 \; Kbytes $ & $ 10 $ \\ 
		 \midrule[0pt]
		\rowcolor{row2} & $ 13 \; Kbytes $ & $ 9 $ \\ 
		%\cline{2-3}
		\rowcolor{row2}\multirow{2}{*}[6mm]{$ 352\times288 \; pixeles $} & $ 4 \; Kbytes $ & $ 3 $ \\ 
		\bottomrule
	\end{tabular} }
	\caption{Cantidad de paquetes observados por ráfaga, dependiendo del nivel de compresión de la cámara.}
	\label{table:camera_packets}
\end{table}

\section{Análisis de la pérdida de paquetes}

Como ocurre en un escenario real, los flujos no se inician al mismo tiempo; por ello, para las simulaciones se ha incluido un período de inicio en el cual todos los flujos comienzan de manera aleatoria. Inicialmente se realiza una serie de pruebas preliminares con $ 100 $, $ 50 $ y $ 40 $ realizaciones, cuyo objetivo es determinar la cantidad de repeticiones necesarias para las pruebas con un adecuado consumo de recursos. Los valores medios de la pérdida de paquetes se calcularon para cada caso considerando su respectiva cantidad de repeticiones ($ 100 $, $ 50 $ y $ 40 $), para dichas pruebas los valores fueron muy similares en los tres casos. Por este motivo, cada prueba se repite $ 40 $ veces y los resultados que se muestran son los correspondientes a los valores medios de dichos resultados. Además, cada valor incluye un intervalo de confianza del $ 95 \% $ que se muestra en los gráficos. La duración de cada prueba es de $ 60 \; s $, tiempo que asegura un patrón de tráfico estable para los flujos.

Los resultados de la pérdida de paquetes cuando se ha utilizado el tráfico de una sola cámara (lo que equivale aproximadamente a un $ 29 \% $ de la utilización del enlace) es cero para todas las pruebas con diferentes tamaños de \textit{buffer} (desde $ 30 $ hasta $ 65 $ paquetes) y para diferentes capacidades de la red interna ($ 10 \; Mbps $ y $ 100 \; Mbps $). Esto se debe a que el tamaño de la ráfaga ($ 26 $ paquetes) es menor que el tamaño del \textit{buffer} en todos los casos, y por lo tanto, el \textit{buffer} puede absorber todos los paquetes entrantes sin producir pérdidas que deterioren la calidad de la comunicación.

Para los casos en los que se utilizan dos y tres flujos de datos de cámara el ancho de banda disponible es de $ 57 \% $ y $ 85 \% $ respectivamente. En estos casos la pérdida de paquetes puede ser inaceptable como se muestra en las Figuras \ref{fig:camera_2} y \ref{fig:camera_3} (notar que los ejes ``\textit{y}'', tienen diferentes escalas para cada figura). La causa es el solapamiento de las ráfagas que provienen de diferentes cámaras, ya que producen una ráfaga de tráfico que puede exceder la capacidad del \textit{buffer}. Como ejemplo se destacan los resultados deficientes que se obtienen cuando se utiliza un tamaño de \textit{buffer} de $ 30 $  paquetes y dos flujos de cámaras, en dicho caso, la pérdida de paquetes casi alcanza un $ 4 \% $ y un $ 9 \% $ para capacidades de la red interna de $ 10 \; Mbps $ y $ 100 \; Mbps $ respectivamente. Si el número de paquetes generados por cada cámara en una ráfaga es de $ 26 $ paquetes, es fácil que el \textit{buffer} se llene cuando diversas ráfagas llegan.

\InsertFig{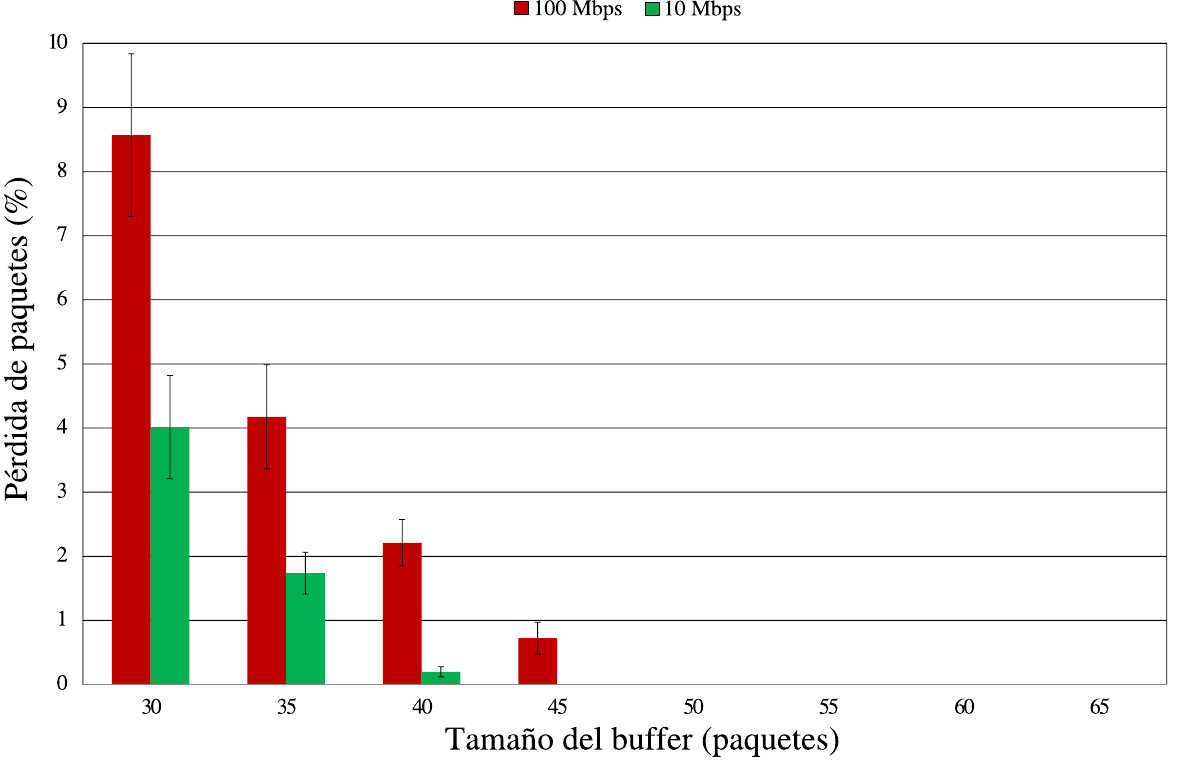}{fig:camera_2}{Relación entre el tamaño del \textit{buffer} y la pérdida de paquetes para dos flujos de cámara IP.}{}{0.9}{height=7cm}{}

\InsertFig{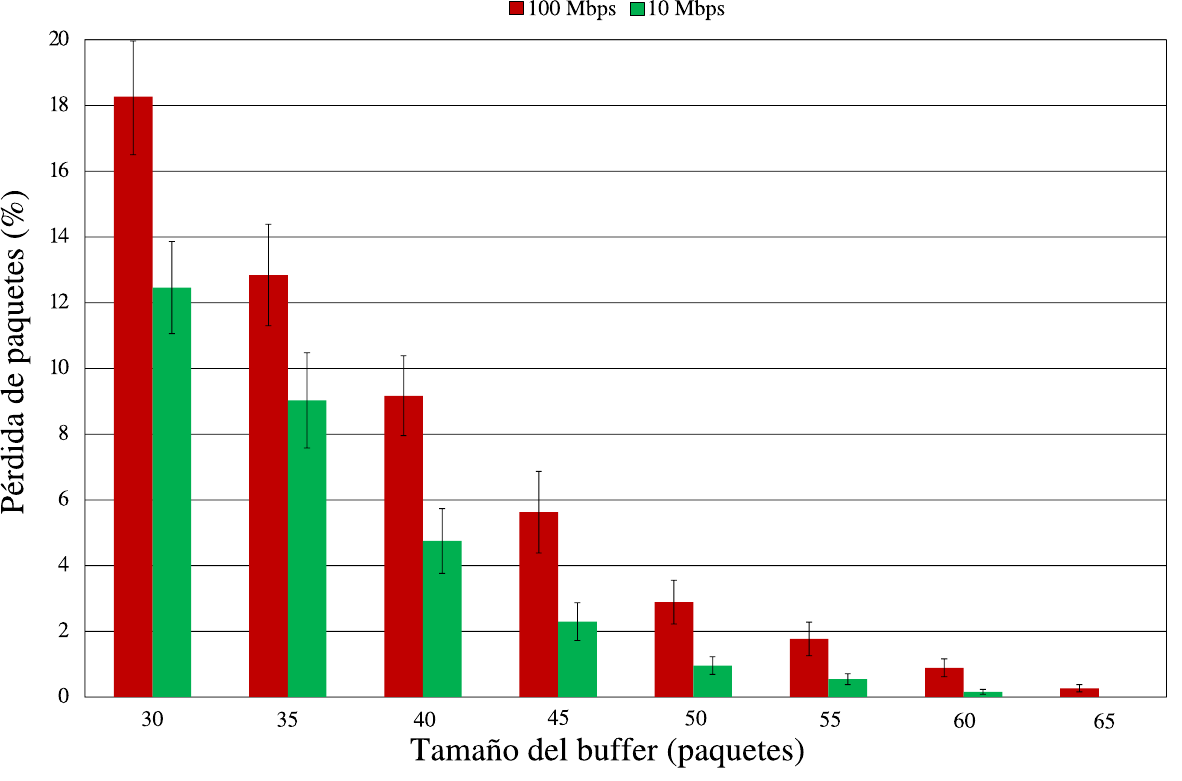}{fig:camera_3}{Relación entre el tamaño del \textit{buffer} y la pérdida de paquetes para tres flujos de cámara IP.}{}{0.9}{height=7cm}{}

Al mismo tiempo, otro fenómeno interesante se puede observar cuando se comparan los resultados para $ 10 \; Mbps $ y $ 100 \; Mbps $ en ambas figuras. Se puede observar que la pérdida de paquetes es más alta cuando la capacidad de la red es de $ 100 \; Mbps $. Además, hay algunos casos en los que la pérdida de paquetes solamente aparece para la red más rápida, como sucede en la Figura \ref{fig:camera_2} para un \textit{buffer} de $ 45 $ paquetes. Esto sucede porque la velocidad de la conexión a Internet sigue siendo la misma, y entonces, cuando una ráfaga es generada por la cámara en una red de $ 100 \; Mbps $ el \textit{buffer} se llenará más rápidamente que en una red de $ 10 \; Mbps $. En estos casos, el incrementar la capacidad de la red de $ 10 \; Mbps $ a $ 100 \; Mbps $ producirá ráfagas de paquetes perdidos, degradando el rendimiento de la red.

\section{Distribución de la pérdida de paquetes}

El análisis anterior ha permitido mostrar la relación que existe entre el comportamiento del tráfico a ráfagas y la pérdida de paquetes y por consiguiente la \gls{qos}. También, resulta interesante analizar algunos detalles del comportamiento de dicho tráfico, ya que parece haber una distribución no uniforme de los datos, para ello se han seleccionado los resultados correspondientes a un \textit{buffer} con tamaño de $ 40 $ paquetes en una red de $ 100 \; Mbps $. Las Figuras \ref{fig:histograma_camara_2} y \ref{fig:histograma_camara_3} muestran un histograma de la pérdida de paquetes para las pruebas con dos y tres flujos de cámara respectivamente. En dichos histogramas se puede observar el porcentaje de las iteraciones que han alcanzado un determinado valor de pérdidas. 

\InsertFig{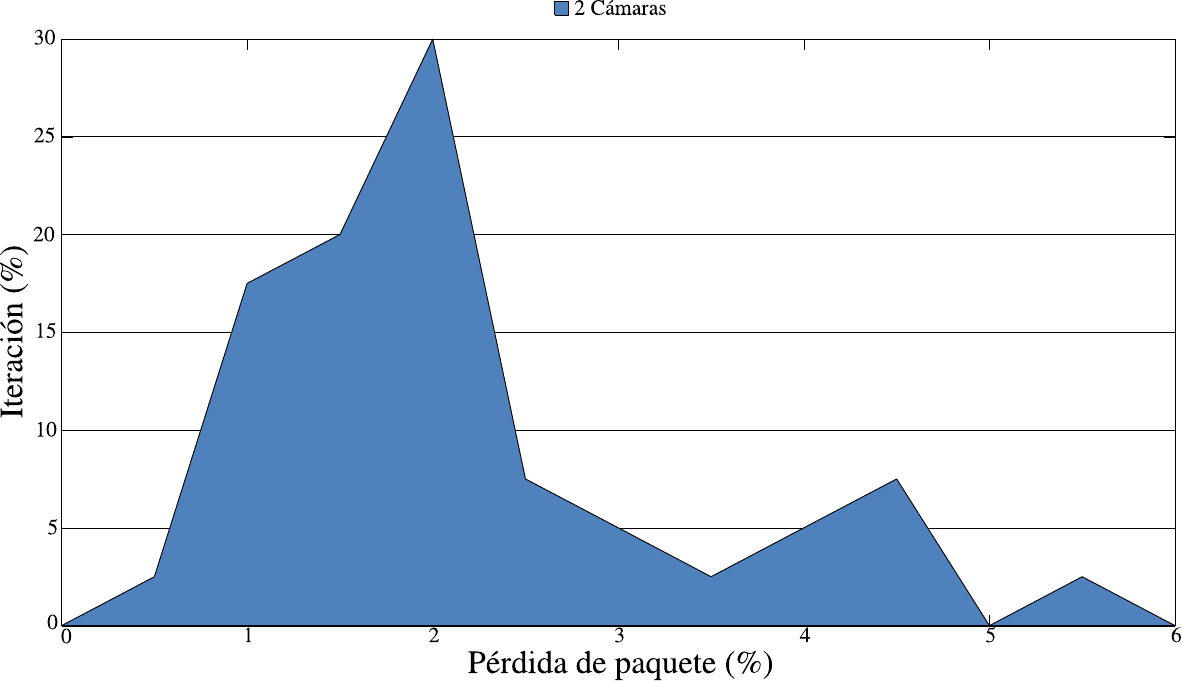}{fig:histograma_camara_2}{Pérdida de paquetes para dos flujos de cámara IP que atraviesan un \textit{buffer} de $ 40 $ paquetes en una red de $ 100 \; Mbps $.}{}{0.9}{height=7cm}{}

\InsertFig{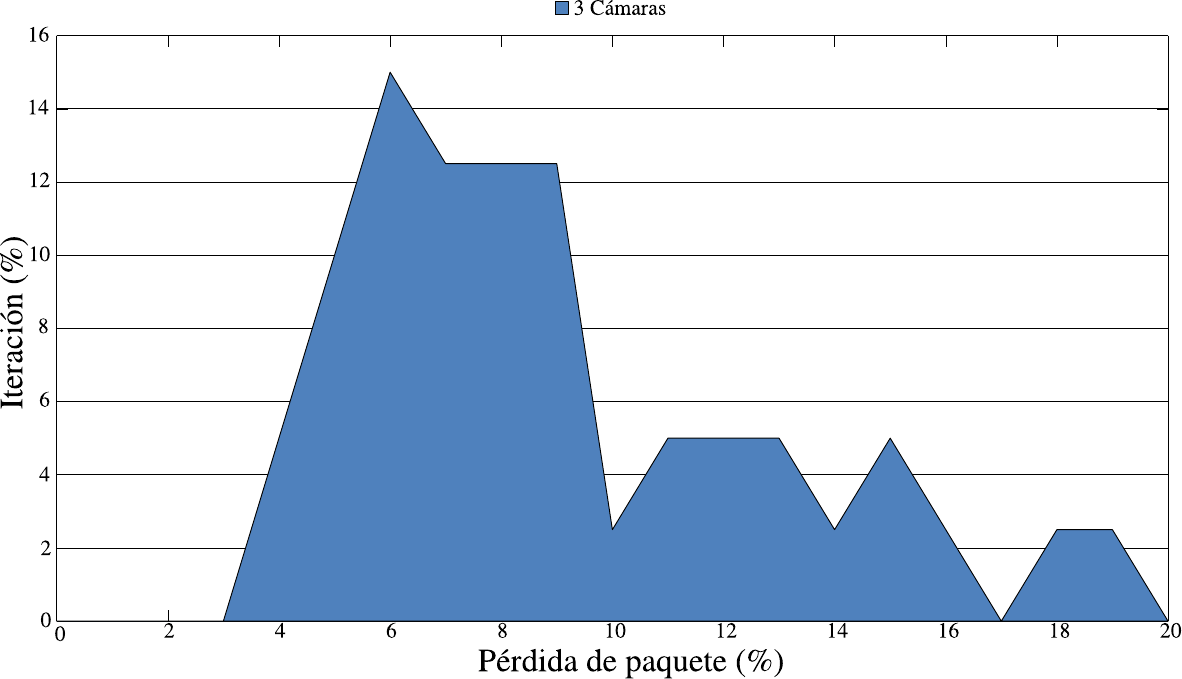}{fig:histograma_camara_3}{Pérdida de paquetes para tres flujos de cámara IP que atraviesan un \textit{buffer} de $ 40 $ paquetes en una red de $ 100 \; Mbps $.}{}{}{height=7cm}{}

En ambos casos los datos muestran una distribución no uniforme, presentándose algunas variaciones entre los valores obtenidos de la pérdida de paquetes para las diferentes repeticiones de una misma prueba. Por ejemplo, en la Figura \ref{fig:histograma_camara_2} se muestra que algunas de las pruebas tienen un $ 0 \% $ de pérdidas mientras que en otros casos dicho valor supera el $ 5 \% $. En la Figura \ref{fig:histograma_camara_3} un alto porcentaje de las pruebas ha obtenido una pérdida de paquetes superior al $ 10 \% $. Estos resultados sugieren que bajo las mismas condiciones de la red, la \gls{qos} de algunas comunicaciones de este tipo de servicio podrían verse drásticamente degradadas. %Estos resultados deben ser considerados como datos preliminares y como una iniciativa para futuras líneas de investigación.

\chapter{El \textit{Buffer} y las aplicaciones}
\label{cha:Buffer y applicaciones}
% \minitoc
En este capítulo se realiza un análisis del efecto del tamaño del \textit{buffer} en la presencia de tráfico a ráfagas y sus posibles implicaciones en el tráfico de otras aplicaciones que comparten un enlace en común. Se ha seleccionado como ejemplo de análisis un entorno de \gls{pymes} (pero igualmente aplicable al caso de una persona trabajando remotamente, por ejemplo desde la casa) con un solo enlace de acceso hacia Internet, en el cual convergen servicios de \gls{voip}, videoconferencia y videovigilancia. En este escenario se realizan dos pruebas principales: 

\begin{itemize}
    \item En la primera, se valora el efecto de la variación del tamaño del \textit{buffer} cuando la utilización del enlace se mantiene fija. 
    \item En la segunda, se observan los efectos del cambio de la utilización del enlace para un determinado tamaño de \textit{buffer}.
\end{itemize}

Además, se analiza el efecto del aumento de la capacidad de la red interna (como se realizó en al capítulo \ref{cha:Buffer y rafagas}) cuando convergen los servicios mencionados. Dados los resultados obtenidos en el capítulo \ref{cha:Buffer y rafagas}, en este capítulo se describen las distribuciones de pérdida de paquetes por medio de histogramas, ya que la mayoría de los resultados presentan un buen nivel de \gls{qos}, y sin embargo, unos pocos presentan peores niveles. El análisis de calidad está basado principalmente en dos parámetros: pérdida de paquetes por flujo y retardo. Además, para el caso de \gls{voip}, también se presentan resultados utilizando estimadores subjetivos de la calidad basados en estos parámetros objetivos, utilizando diferentes valores de retardo de red. Por otro lado, se ha considerado el desbordamiento del \textit{buffer} como la única causa de pérdida de paquetes.    

\section{Escenario de red propuesto}

En la Figura \ref{fig:test_40_70_v2} se muestra el escenario utilizado para las pruebas, el cual consiste en un enlace de acceso a Internet donde convergen dos flujos generados por cámaras \gls{ip} con un ancho de banda de $ 1 \; Mbps $ cada uno, una sesión de videoconferencia con un ancho de banda medio de $ 1.5 \; Mbps $ y dos llamadas de \gls{voip} con un ancho de banda de $ 24 \; Kbps $ cada una, lo que supone un total de ancho de banda generado de $ 3.5 \; Mbps $. Para este escenario se plantean dos pruebas diferentes:

\begin{itemize}
    \item En la primera, la capacidad del enlace de acceso a Internet se establece en $ 5 \; Mbps $, de esta manera la utilización media del enlace se fija al $ 70 \% $ y se realizan las pruebas para diferentes tamaños de \textit{buffer}.
    \item En la segunda prueba, el tamaño del \textit{buffer} del \textit{router} de acceso a Internet se fija en $ 40 $ paquetes y las simulaciones se repiten para diferentes valores de la capacidad de acceso, y por consiguiente, para diferentes niveles de la utilización del enlace, dentro de un rango que va desde el $ 50 \% $ al $ 90 \% $.
\end{itemize}

\InsertFig{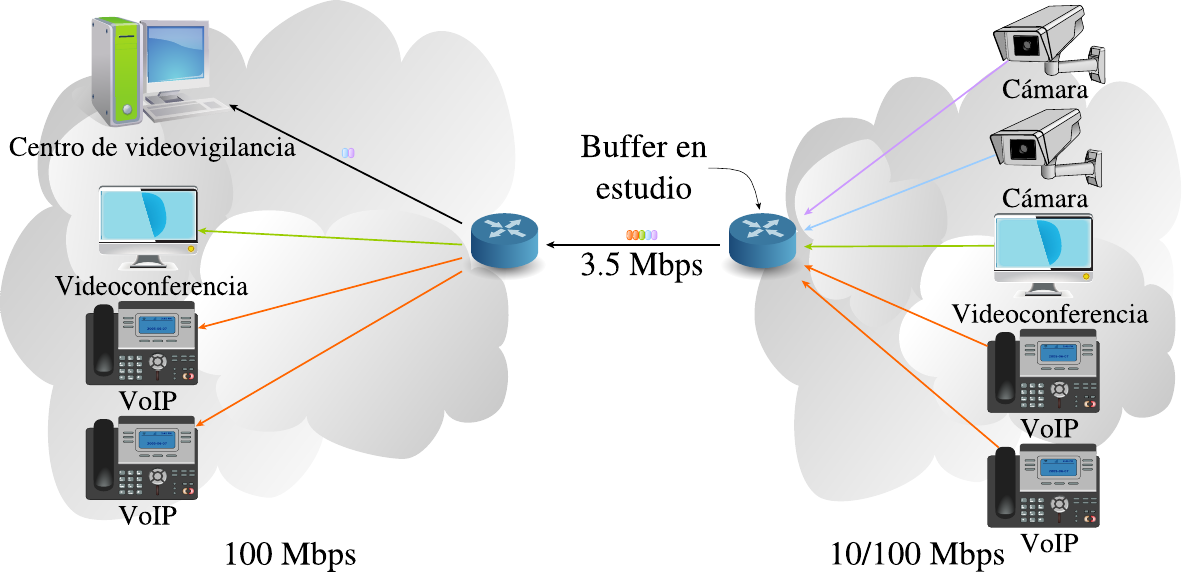}{fig:test_40_70_v2}{Escenario para las pruebas con dos conexiones de cámaras, una videoconferencia y dos llamadas de VoIP.}{}{0.9}{}{}

\section{Tráfico utilizado}

Con la finalidad de desarrollar las pruebas descritas anteriormente, se han utilizado tres fuentes de tráfico multimedia diferentes: videovigilangia, videoconferencia y \gls{voip}. Para el tráfico de videovigilancia y videoconferencia no se utilizan modelos de tráfico, sino, trazas de tráfico real que fueron capturadas previamente en escenarios reales para luego ser generadas en \gls{ns}, usando sus tamaños de paquetes y tiempo entre paquetes. El tráfico de VoIP es generado mediante un agente \gls{cbr} de \gls{ns}. 

La metodología utilizada para las capturas del tráfico de videoconferencia se ilustra en la Figura \ref{fig:vidyo}. Para dichas trazas se ha utilizado la arquitectura de Vidyo\texttrademark, la cual incorpora la tecnología \gls{avl} que permite la optimización dinámica del video para cada terminal, aprovechando la tecnología de compresión $ H$.$264 $-\gls{svc}. La aplicación de videoconferencia se configuró con $ 2 \; Mbps $ de ancho de banda (sin embargo, la captura real de dicho tráfico solo alcanza un ancho de banda de $ 1.5 \; Mbps $) y una resolución de $ 800 \times 450 \; px $ mientras la cámara capturaba un video con mucho movimiento (un partido de fútbol). 

\InsertFig{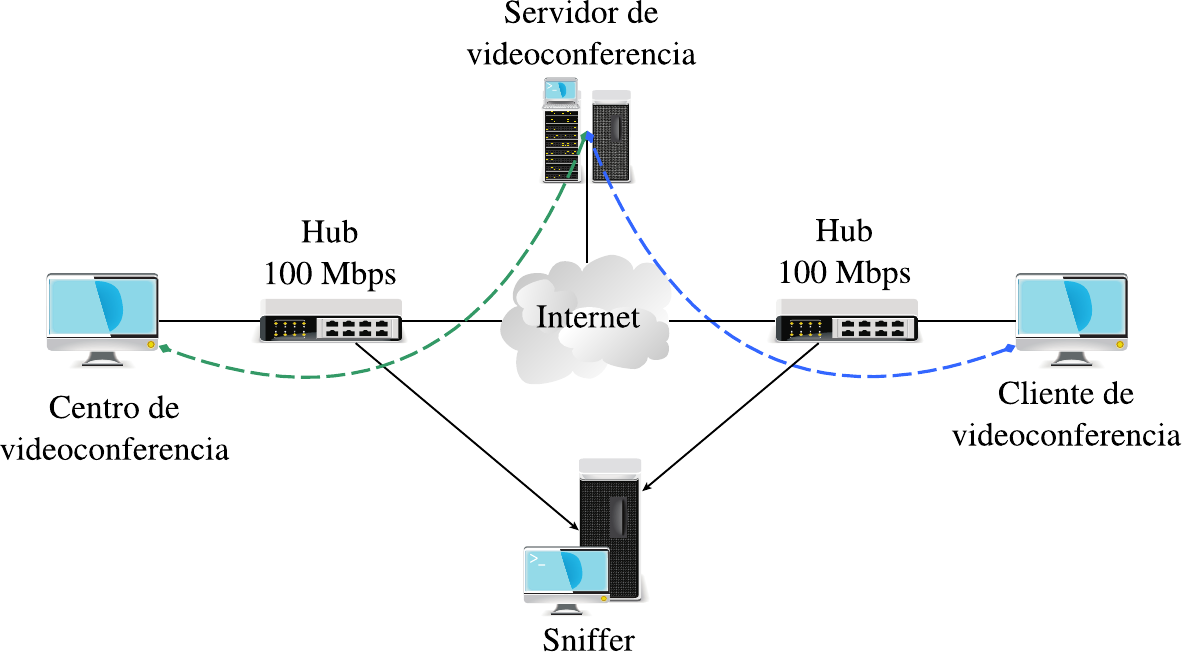}{fig:vidyo}{Escenario para la captura del tráfico de una videoconferencia.}{}{0.9}{}{}

El tráfico de voz se genera de acuerdo a la recomendación $ G $.$ 729 $ con un tiempo entre paquetes de $ 20 \; ms $ y $ 2 $ muestras por paquete, resultando en un tamaño de paquete de $ 60 \; bytes $. Para las trazas de videovigilancia, se han utilizado las mismas que se usaron para obtener los resultados del capítulo \ref{cha:Buffer y rafagas}.

\section{Análisis de pérdida de paquetes}

Este análisis se enfoca en la calidad que se obtiene para el tráfico combinado y para cada uno de los servicios que comparten la red, lo cual corresponde a un estudio más detallado del comportamiento del tráfico en el escenario en cuestión. Además, el  análisis del tráfico combinado se centra en la relación de pérdida de paquetes para los casos en los que la red interna es de $ 10 \; Mbps $ y $ 100 \; Mbps $, mientras que el análisis de los flujos lo hace en el caso de una red interna a $ 100 \; Mbps $, ya que esta capacidad de red es la que presenta el peor de los casos en términos de pérdida de paquetes. 

\subsection{Pérdida de paquetes del tráfico combinado}

Los resultados para el tráfico combinado de los tres tipos de flujo que comparten la red, correspondientes a la primera prueba, se muestran en la Figura 	\ref{fig:scenario_70_total}. En la cual se observa la pérdida de paquetes para diferentes tamaños de \textit{buffer} cuando la utilización del enlace es del $ 70 \% $ con sus respectivos intervalos de confianza del $ 95 \% $. En dicho gráfico se puede observar el mismo fenómeno mencionado en el capítulo \ref{cha:Buffer y rafagas}: la pérdida de paquetes es mayor cuando la capacidad de la red local es de $ 100 \; Mbps $. Además, dicho efecto aumenta cuando el tamaño del \textit{buffer} disminuye.

\InsertFig{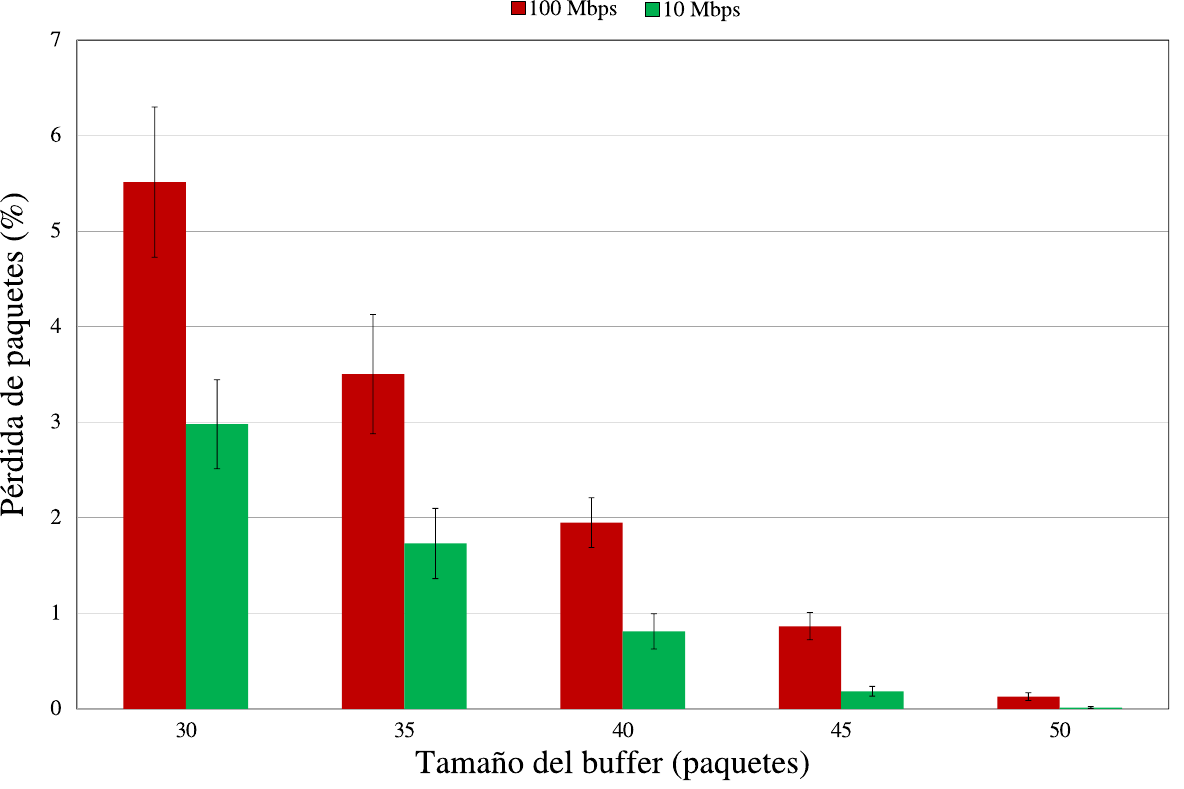}{fig:scenario_70_total}{Relación entre el tamaño del \textit{buffer} y la pérdida de paquetes para una utilización del enlace del $ 70 \% $.}{}{0.9}{height=7cm}{}

A pesar de que el tráfico no se muestra de manera separada para cada flujo (este tema se analiza en la siguiente sección), la pérdida de paquetes afecta a todas las aplicaciones, de esta manera se observa que la presencia de aplicaciones que generan tráfico a ráfagas (videovigilancia) causa la pérdida de paquetes para todas las aplicaciones que coexisten, incluso para aquellas que generan tráfico a una tasa de $ bit $ constante (\gls{voip}).

Para la segunda prueba, los resultados se muestran en la Figura \ref{fig:scenario_40_total}, la cual describe la relación de la pérdida de paquetes y la utilización del enlace para este escenario en particular. Como era de esperar, la pérdida de paquetes se incrementa cuando la utilización del enlace crece para el caso de un tamaño de \textit{buffer} de $ 40 $ paquetes. De nuevo, la pérdida de paquetes es mayor cuando la capacidad de la red es de $ 100 \; Mbps $.

\InsertFig{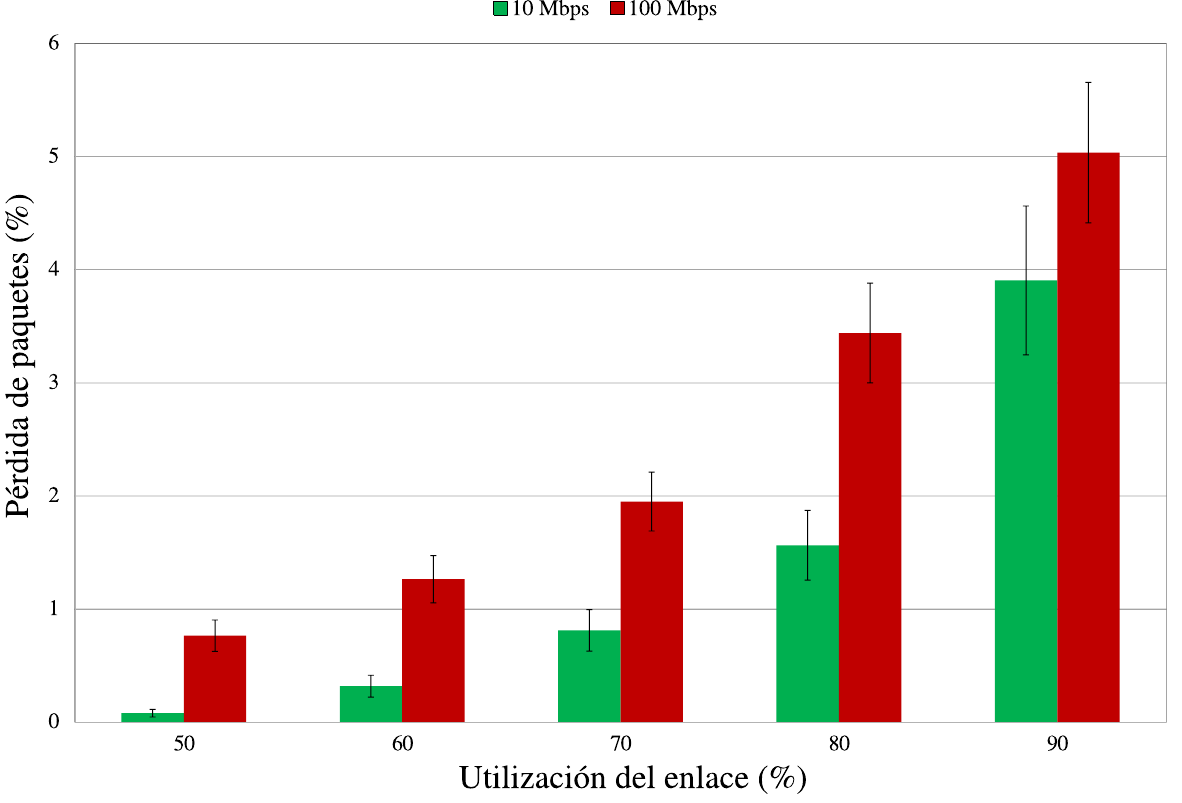}{fig:scenario_40_total}{Relación entre la utilización del enlace y la pérdida de paquetes para un tamaño de \textit{buffer} de $ 40 $ paquetes.}{}{0.9}{height=7cm}{}

\subsection{Pérdida de paquetes por flujo}

Para este caso se han desarrollado dos tipos diferentes de pruebas: en la primera se considera un escenario con la utilización de enlace fijada y se varía el tamaño del \textit{buffer}, en la segunda se fija el tamaño del \textit{buffer} variando la utilización del enlace. Las Figuras \ref{fig:scenario_70_flows} y \ref{fig:scenario_70_flows_percent} muestran la pérdida de paquetes por flujo usando una utilización del enlace fija ($ 70 \% $) con sus correspondientes intervalos de confianza del $ 95 \% $. 

\InsertFig{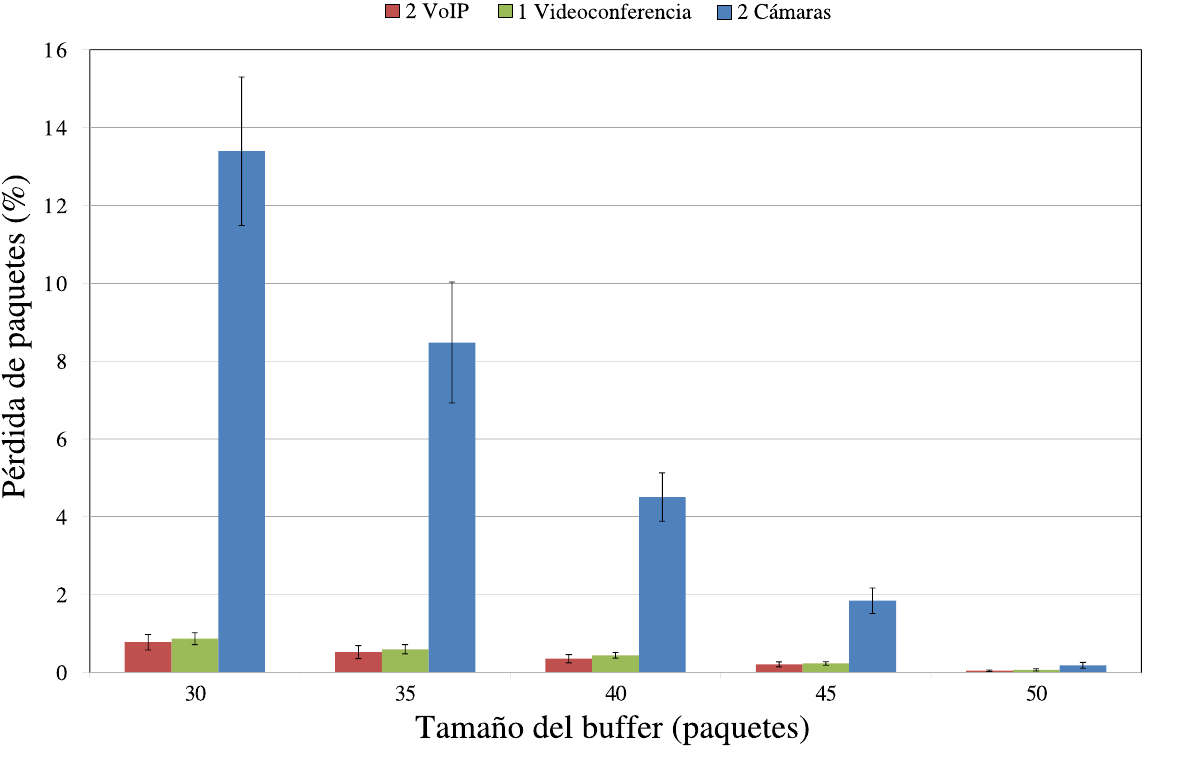}{fig:scenario_70_flows}{Pérdida de paquetes por flujo cuando la utilización del enlace es del $ 70 \% $ para diferentes tamaños de \textit{buffer}.}{}{0.9}{height=7cm}{}

\InsertFig{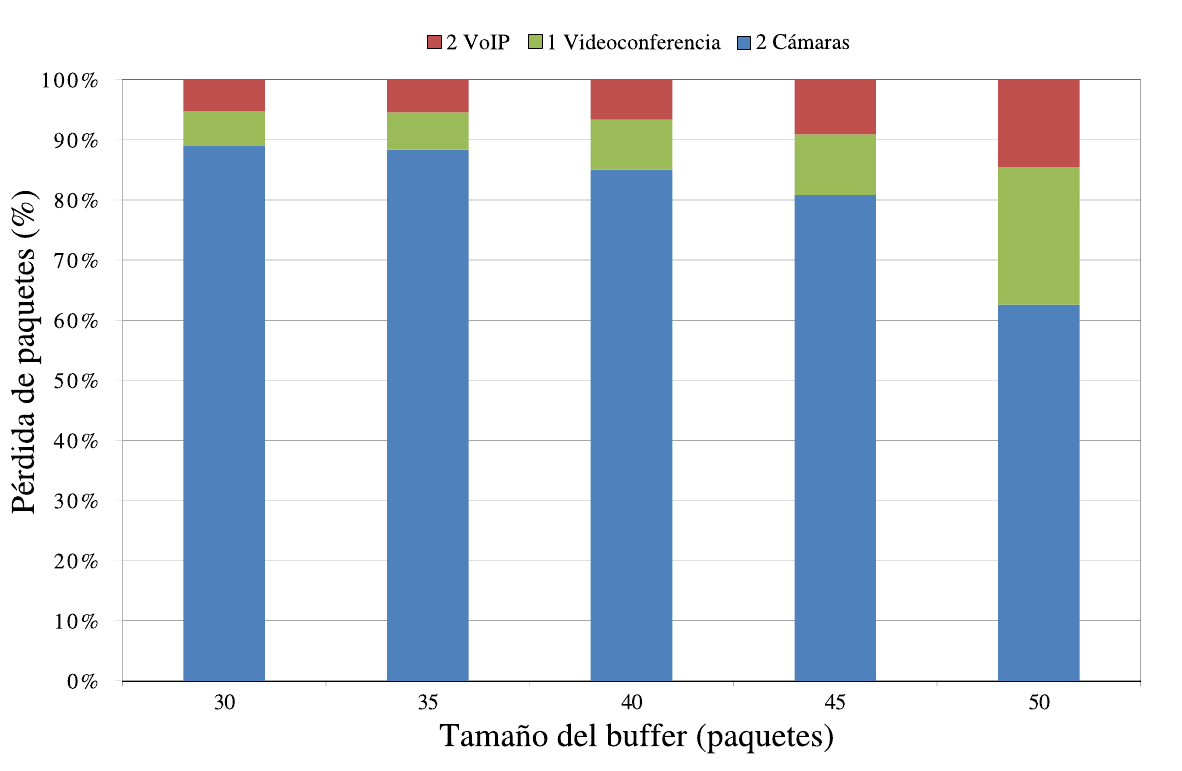}{fig:scenario_70_flows_percent}{Distribución por flujo de la pérdida de paquetes cuando la utilización del enlace es del $ 70 \% $ para diferentes tamaños de \textit{buffer}.}{}{0.9}{height=7cm}{}

La principal causa de pérdida de paquetes es la presencia de una aplicación  que genera tráfico a ráfagas (videovigilancia), la cual causa un desbordamiento del \textit{buffer} y degrada la calidad de todas las aplicaciones coexistentes. Se puede observar que la pérdida de paquetes decrece cuando el tamaño del \textit{buffer} se incrementa, porque los \textit{buffer} más grandes pueden absorber mejor las ráfagas producidas por el tráfico mezclado. Sin embargo, esto podría incrementar el tiempo que un paquete está encolado durante períodos de congestión, generando un mayor retardo. No obstante, la videoconferencia y \gls{voip} obtienen mejores resultados debido a que su perfil de tráfico tiene menos ráfagas. 

Por otro lado, en la Figura \ref{fig:scenario_70_flows_percent} se puede ver que la distribución de la pérdida de paquetes no es la misma para todas las pruebas con diferentes tamaños de \textit{buffer}. Los \textit{buffer} pequeños aumentan el problema causado por el tráfico de videovigilancia (el que presenta más ráfagas), incrementando la tasa de paquetes perdidos correspondientemente a este servicio.

Las Figuras \ref{fig:scenario_40_flows} y \ref{fig:scenario_40_flows_percent} muestran los resultados de las pruebas con el tamaño del \textit{buffer} fijo ($ 40 $ paquetes). En los gráficos se representa en el eje ``\textit{x}'', la utilización media del enlace de acuerdo al ancho de banda generado por las aplicaciones. Como era de esperar, la pérdida de paquetes se incrementa cuando la utilización del enlace crece. De nuevo, la distribución de la pérdida de paquetes no es la misma para todas las pruebas (Figura \ref{fig:scenario_40_flows_percent}), aunque las diferencias no son significativas. Así, teniendo en cuenta los resultados de las Figuras \ref{fig:scenario_40_flows} y \ref{fig:scenario_40_flows_percent}, se puede ver que el tamaño del \textit{buffer} tiene una fuerte influencia en la distribución de la pérdida de paquetes por flujo. 

\InsertFig{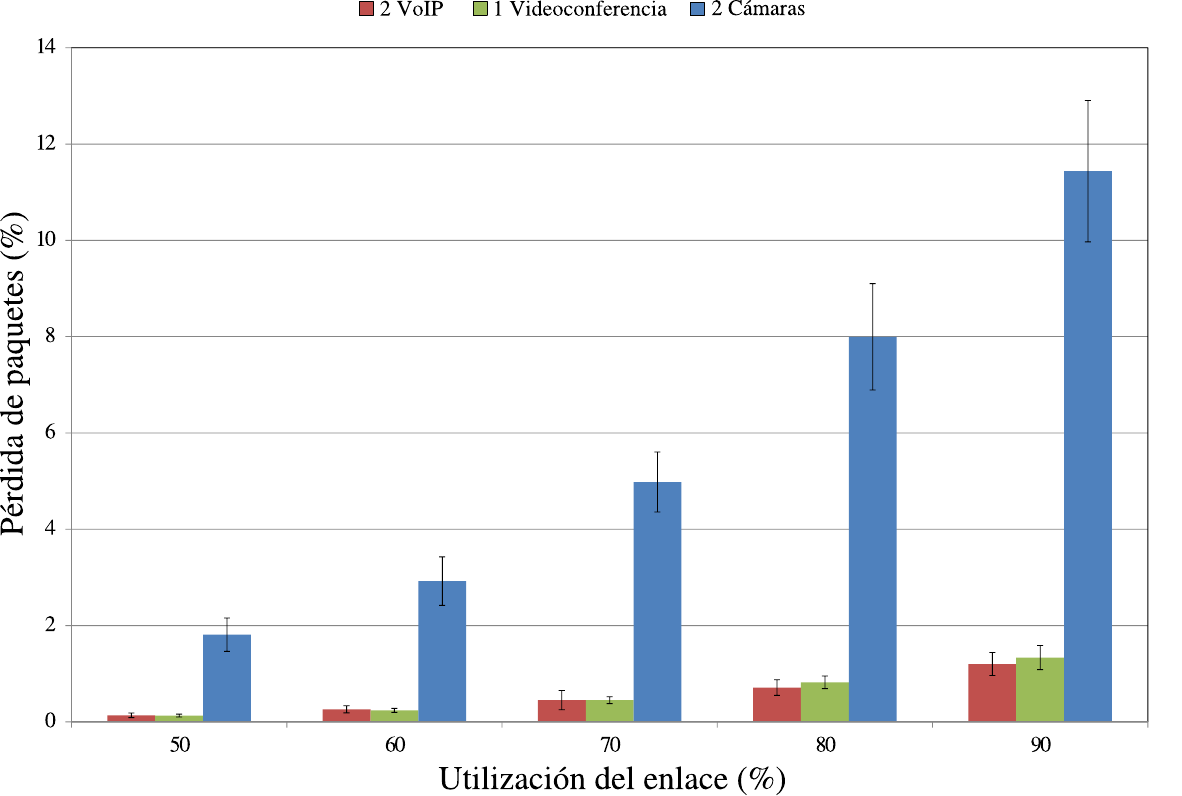}{fig:scenario_40_flows}{Relación entre la pérdida de paquetes por flujo y la utilización del enlace para un \textit{buffer} de $ 40 $ paquetes.}{}{0.9}{height=7cm}{}

\InsertFig{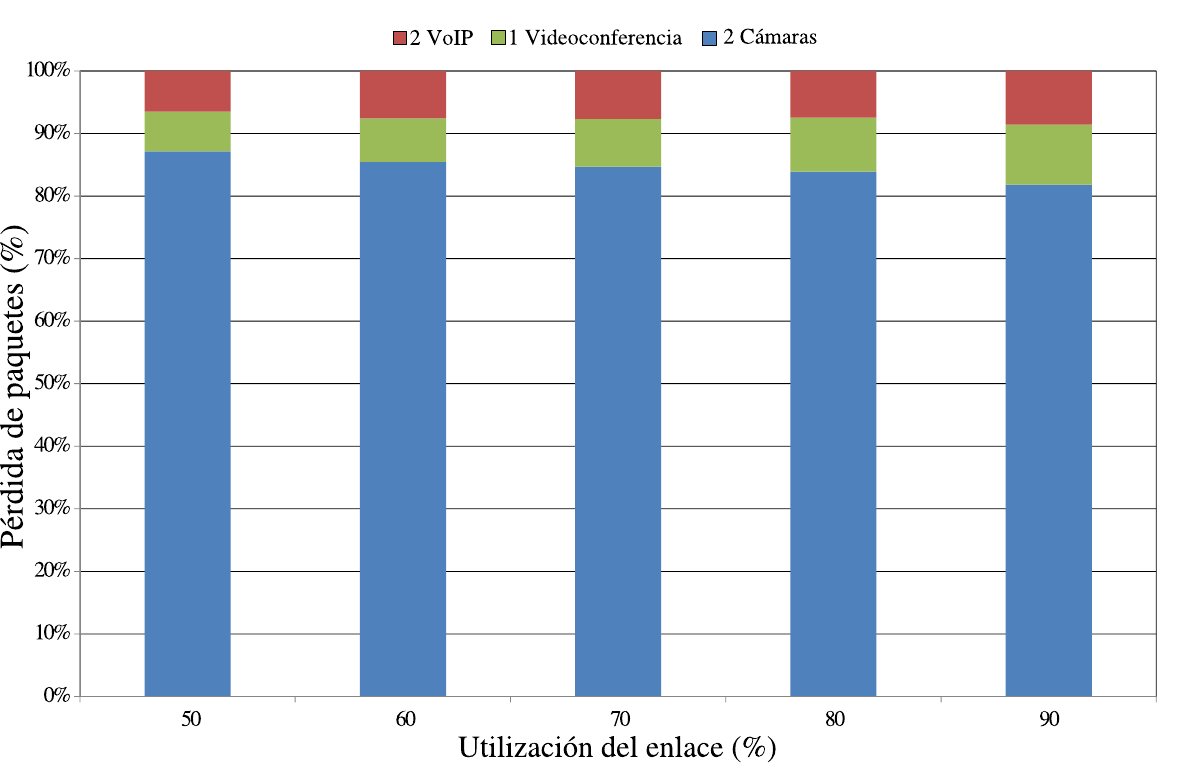}{fig:scenario_40_flows_percent}{Distribución por flujo de la pérdida de paquetes para un \textit{buffer} de $ 40 $ paquetes en función de la utilización del enlace.}{}{0.9}{height=7cm}{}

\section{Distribución de la pérdida de paquetes}

En la sección anterior se han presentado la media de los resultados para una serie de pruebas y se han obtenido valores pequeños de los intervalos de confianza. Sin embargo es necesario describir la distribución de pérdida de paquetes entre las comunicaciones establecidas en las diferentes pruebas, ya que la pérdida de paquetes podría no ser uniforme entre ellas, como se ha comentado en el capítulo \ref{cha:Buffer y rafagas}. Mientras que en algunas pruebas no se pierden paquetes, en otras, algunos flujos presentan altas tasas de paquetes perdidos, porque en esos casos el solapamiento de los flujos es más grande. Como resultado, en la misma red habrá momentos en que con las mismas condiciones, una comunicación puede obtener una muy buena calidad mientras que en otros la calidad presenta valores significativamente peores. La principal causa de este efecto es la distribución aleatoria de las superposiciones entre las ráfagas.

A continuación se introduce una forma de medir este fenómeno. Para esto, se ha seleccionado un escenario con una utilización del enlace del $ 70 \% $ y un tamaño de \textit{buffer} de $ 40 $ paquetes, y las mismas aplicaciones descritas anteriormente. En este escenario específico, las pruebas se han repetido $ 200 $ veces (a pesar que en el capítulo \ref{cha:Buffer y rafagas} se mencionó que con $ 40 $ repeticiones se obtienen resultados muy similares) para observar mejor el solapamiento de los flujos y su relación con la pérdida de paquetes. Los resultados se presentan por medio de un histograma correspondiente al tráfico total y para cada servicio donde en el eje ``\textit{x}'', se  muestra el porcentaje de pérdida  de paquetes, y en el eje ``\textit{y}'', el porcentaje de iteraciones en la cual se ha obtenido ese valor de pérdida de paquetes.

En la Figura \ref{fig:histograma_total} se presenta un histograma de la pérdida de paquetes para el tráfico total en una red a $ 100 \; Mbps $. El valor medio de la pérdida de paquetes correspondiente a los resultados  de las $ 200 $ repeticiones es de $ 2.11 \% $. Como se puede observar en dicha figura, existe una gran cantidad de iteraciones que se encuentran por debajo de la media, incluso el $ 1 \% $ de los casos sin pérdida de paquetes, a la vez, muchas de las pruebas duplican la media y algunas se encuentran por encima del $ 5 \% $.

\InsertFig{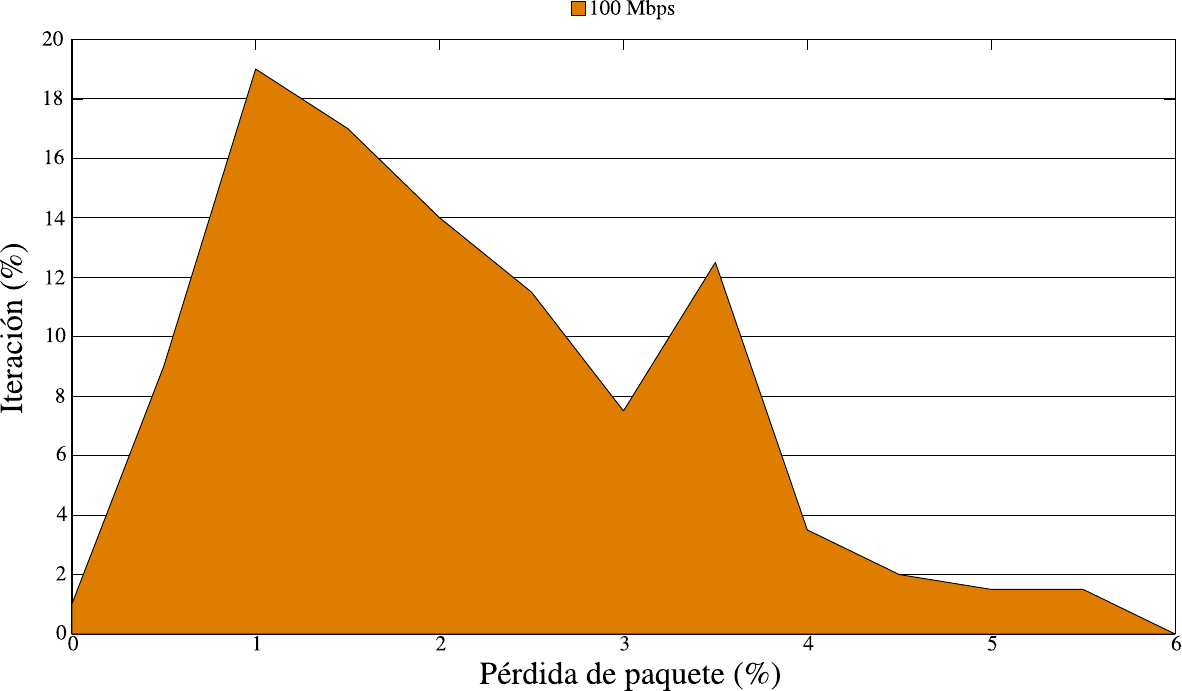}{fig:histograma_total}{Pérdida de paquetes para el tráfico combinado con un \textit{buffer} de $ 40 $ paquetes y una utilización del enlace del $ 70 \% $.}{}{0.9}{height=7cm}{}

El caso de \gls{voip} se presenta en la Figura \ref{fig:histogram_1}, en la cual casi el $ 80 \% $ de las llamadas presentan un valor de pérdida de paquetes menor al $ 0.75 \% $. La pérdida de paquetes aumenta hasta un $ 3 \% $ o más en el $ 0.5 \% $ de los casos (equivalente a $ 20 $ llamadas) en los cuales la \gls{qos} sería significativamente degradada. Esto confirma que hay un porcentaje de llamadas en las cuales la calidad obtenida no será lo suficientemente buena para los usuarios.

\InsertFig{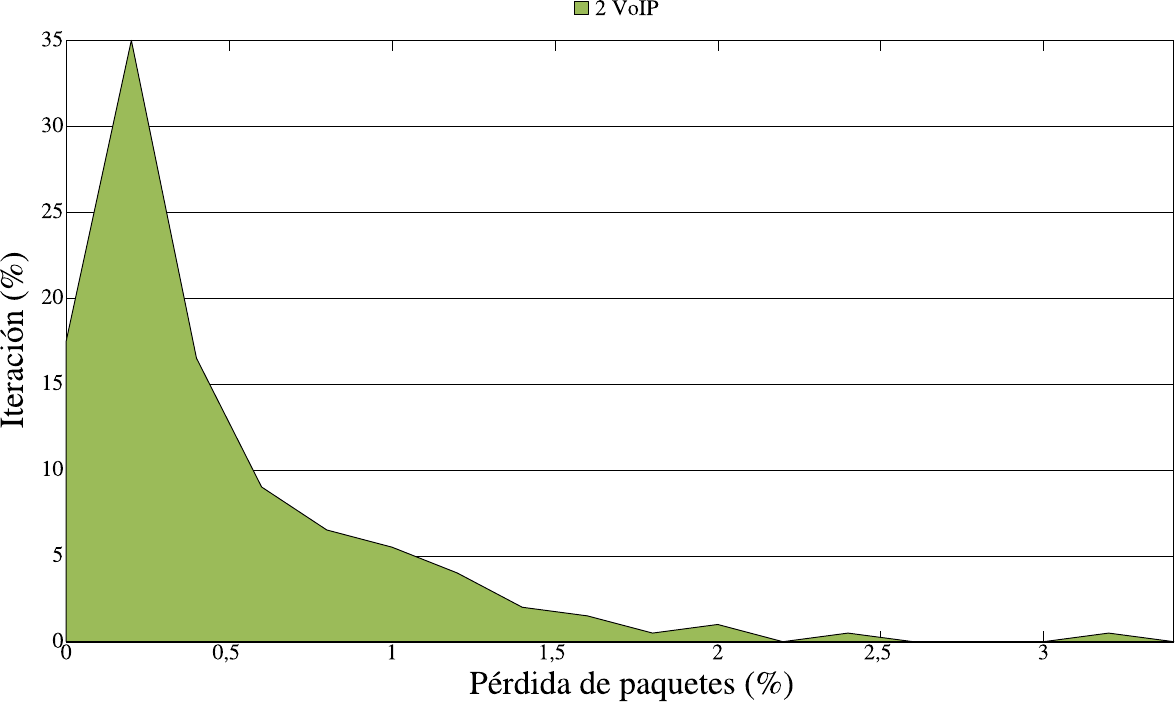}{fig:histogram_1}{Pérdida de paquetes para el tráfico de VoIP con un \textit{buffer} de $ 40 $ paquetes y una utilización del enlace del $ 70 \% $.}{}{0.9}{height=7cm}{}

El servicio de videoconferencia presenta un comportamiento similar (Figura \ref{fig:histogram_2}). La tasa de pérdida de paquetes es baja para un alto porcentaje de las pruebas. Sin embargo, estas pérdidas pueden afectar a la calidad de la videoconferencia. Por otro lado, los resultados de las comunicaciones del servicio de videovigilancia (Figura \ref{fig:histogram_3}) muestran el nivel más alto de pérdida de paquetes (hasta un $ 14 \% $ en algunos casos), el cual podría degradar significativamente la \gls{qos} de este servicio.

\InsertFig{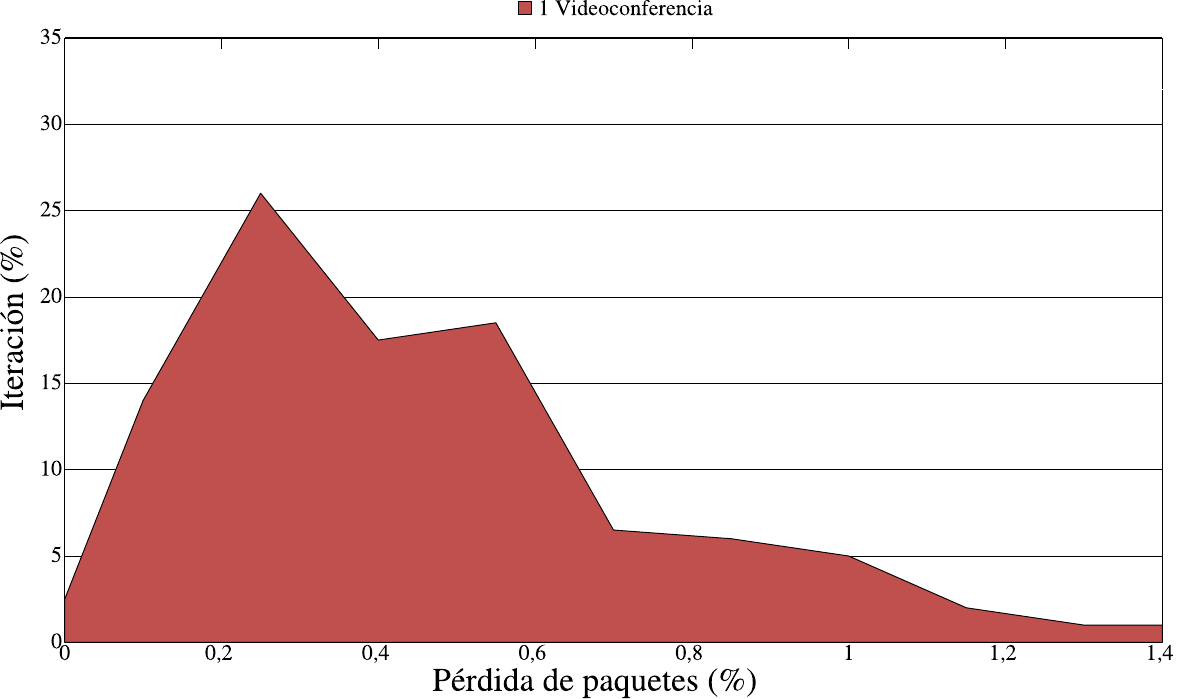}{fig:histogram_2}{Pérdida de paquetes para el tráfico de videoconferencia con un \textit{buffer} de $ 40 $ paquetes y una utilización del enlace del $ 70 \% $.}{}{0.9}{height=7cm}{}

\InsertFig{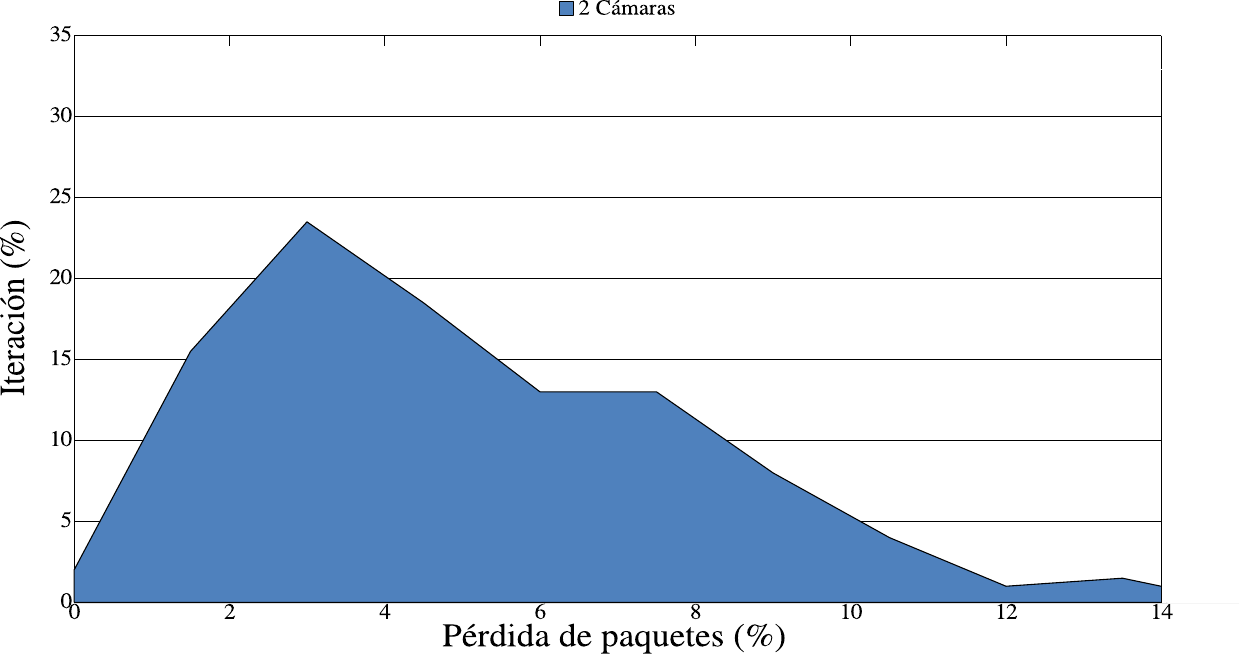}{fig:histogram_3}{Pérdida de paquetes para el tráfico de cámara IP con un \textit{buffer} de $ 40 $ paquetes y una utilización del enlace del $ 70 \% $.}{}{0.9}{height=7cm}{}

\section{MOS para llamadas de VoIP}

Ahora, se analizará el efecto de las ráfagas de pérdidas de paquetes en la calidad subjetiva de \gls{voip}, ya que es un servicio en tiempo real con requerimientos muy específicos de retardo y pérdida de paquetes. En este caso, se han utilizado los resultados del histograma de la pérdida de paquetes para el tráfico de  \gls{voip} analizado anteriormente. Con el objetivo de estimar la calidad subjetiva que se obtendría para cada llamada, se ha calculado el $ R_{factor} $ de acuerdo con \cite{mos1} mediante la siguiente ecuación \ref{eq:r_factor}.

\begin{eqnarray}
\label{eq:r_factor}
R_{factor}&=&94.2 - 0.24 \times delay_{total} - 0.11(delay_{total}-177.3) \nonumber \\ 
&&\times H(delay_{total}-177.3) - 11 \nonumber \\ 
&&- 40 ln(1+(10 \times delay_{total}))
\end{eqnarray}

Donde $ delay_{total} $ es el retardo \gls{owd} y $ H(x) $ es una función escalón. Así, si el retardo se encuentra por debajo de $ 177.3 \; ms $, entonces no afecta al $ R_{factor} $. Sin embargo, si excede este valor, entonces el $ R_{factor} $ sería significativamente menor. Esto responde al fenómeno citado en \cite{mos1}:

\begin{quote}
    \textit{``Para el \gls{owd} menor que $ 177.3 \; ms $, las conversaciones ocurren con normalidad, mientras que cuando el retardo se excede de $ 177.3 \; ms $ la conversación comienza con deformaciones y rupturas; a menudo degenerando en conversaciones tipo \textit{simplex} para los valores de retardo más alto.''}
\end{quote}

A continuación, se obtiene el \gls{mos} a partir del $ R_{factor} $, utilizando la conversión citada en el mismo artículo \cite{mos1}. Para el retardo total se ha considerado incluir el retardo causado por el \textit{buffer} del \textit{router} y la red; además, se ha incluido un \textit{buffer} de \textit{de-jitter} con la finalidad de absorber las variaciones de retardo generadas por el \textit{buffer} del \textit{router}, así los \textit{buffer} del \textit{router} y el de \textit{de-jitter} se compensan mutuamente.

Se han utilizado seis valores diferentes de retardo de red ($ 20 $, $ 40 $, $ 60 $, $ 100 $, $ 120 $ y $ 140 \; ms $) que producen un retardo total de $ 116 $, $ 136 $, $ 156 $, $ 196 $, $ 216 $ y $ 236 \; ms $, respectivamente. Los resultados se presentan por medio de un histograma (Figura \ref{fig:mos}) del \gls{mos} obtenido para cada prueba. Para los tres valores más bajos del retardo de red ($ 20 $, $ 40 $ y $ 60 $), la figura muestra una cantidad significativa de llamadas con una calidad media según el \textit{E-model} de la \gls{itu}-T \cite{emodel, mos1}. Esto representaría malos resultados para los usuarios de \gls{voip}, ya que de este escenario se esperaría que proporcionara la mejor calidad en todos los casos. Además, las colas a la izquierda de la Figura \ref{fig:mos} representan a unas cuantas llamadas con niveles inaceptables de calidad. Por otro lado, para los tres valores de retardo de red más altos ($ 100 $, $ 120 $ y $ 140 \; ms $), en los cuales el retardo total excede el umbral de $ 177.3 \; ms $, se puede observar un comportamiento peor en términos de \gls{mos}. El incremento del retardo de la red produce una reducción significativa en la calidad subjetiva, dando como resultado una calidad baja en algunos casos.

\InsertFig{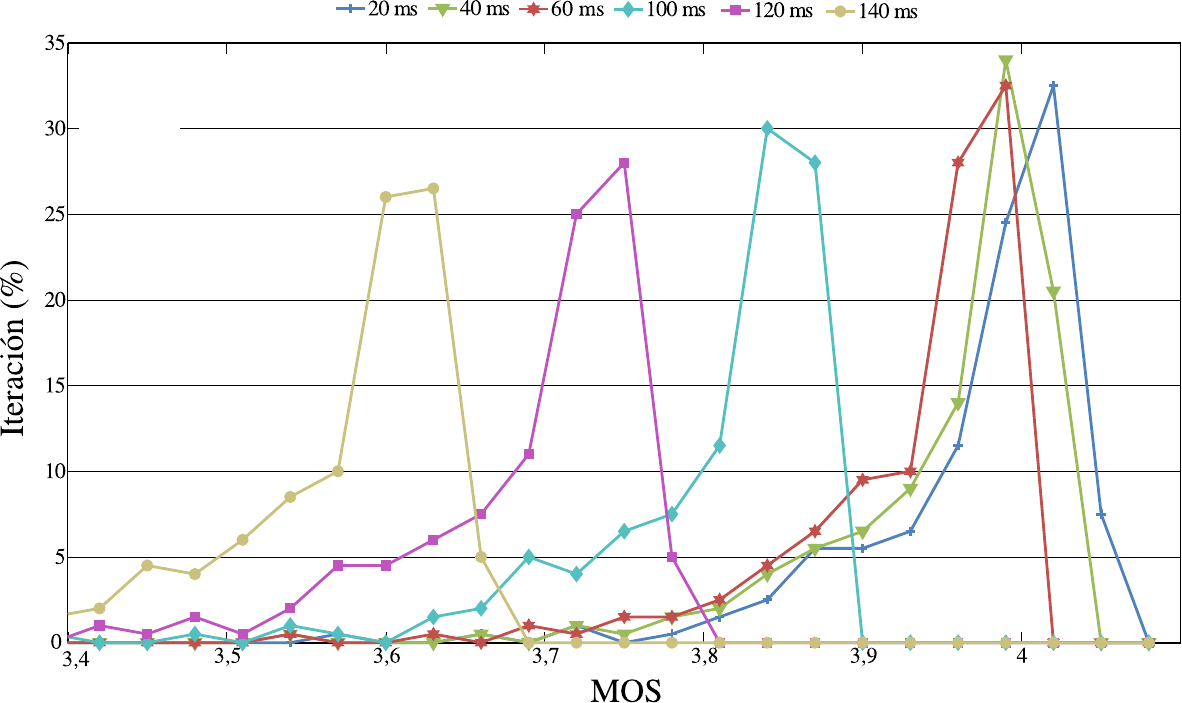}{fig:mos}{MOS con diferentes retardos de red (OWD) para un \textit{buffer} de $ 40 $ paquetes y una utilización del enlace del $ 70 \% $.}{}{0.9}{height=7cm}{}

\chapter{Conclusión}
\label{cha:Conclusion}
% \minitoc
En este capítulo se comentan las principales conclusiones relacionadas con el impacto de los \textit{buffer} en la \gls{qos}, el comportamiento de los flujos de datos y su impacto en la red. También se incluyen aspectos relacionados a la \gls{qos} en la coexistencia de tráficos concurrentes de varios servicios. 

\section{El comportamiento del tráfico}

El comportamiento del tráfico de las aplicaciones está ligado a cómo estas envían datos a la red, y por lo tanto, a las necesidades y requerimientos del servicio asociado a la aplicación y la forma en que dicha aplicación ha sido implementada. En este contexto, algunas aplicaciones generan datos a una tasa constante mientras que en otros casos los patrones de tráfico pueden llegar a ser más complejos, produciendo ráfagas de paquetes que contienen un número de paquetes diferente dependiendo de cada servicio. 

Además, el tamaño de la información juega un papel importante en el tráfico de la red, ya que las aplicaciones generan paquetes de tamaños muy diversos en función de aspectos como: la interactividad, requerimientos temporales y la propia naturaleza de la información, por ejemplo, imágenes de alta calidad y archivos con tamaños muy grandes. El tamaño de los paquetes puede variar desde unas pocas decenas de $ bytes $ en los casos de aplicaciones de \gls{voip} o juegos \textit{online}, hasta el máximo \gls{mtu} que la red permite como en los casos de servicios que necesitan enviar una gran cantidad de información (\gls{iptv}, video \textit{streaming}, entre otros). Muchas de estas aplicaciones incluso generan paquetes de tamaños variados en un mismo flujo, donde un porcentaje de los paquetes son pequeños, y usualmente están relacionados al transporte de información para la gestión y control de la aplicación, mientras que otros son de mayor tamaño y generalmente están asociados al envío de datos propios de la función principal (por ejemplo, video en el caso de sistemas de videoconferencia o videovigilancia).

El tráfico en la red está formado por el conjunto de cada uno de los flujos de información generados por cada una de las aplicaciones utilizadas por los usuarios finales. Por lo tanto, el comportamiento del tráfico en la red es producido por la combinación de todos los tráficos concurrentes, generando patrones de tráfico aún más complejos. En esta combinación de flujos, es muy probable que se formen ráfagas de paquetes por la agrupación aleatoria de diversos flujos, incluso ráfagas aún más grandes cuando coinciden ráfagas de diversas aplicaciones.

\section{La QoS y los \textit{buffer}}

El tamaño del \textit{buffer} se ha identificado como un parámetro crítico a la hora de realizar el planeamiento de una red, principalmente en entornos donde no es posible aumentar la capacidad del enlace de acceso. La razón de esto es la relación entre el tamaño del \textit{buffer} y el número de paquetes contenidos en una ráfaga de tráfico que generan las aplicaciones, ya que dicho número debe ser consistente con la cantidad de paquetes que un \textit{buffer} puede absorber durante períodos de congestión, con la finalidad de prevenir la pérdida de paquetes debido al desbordamiento del \textit{buffer}. Además, se debe tener en cuenta que la tasa de llenado del \textit{buffer} está determinada por la relación entre la velocidad de la red interna y el acceso a Internet, ya que esto podría producir pérdida de paquetes cuando  se envían ráfagas con cantidades de paquetes muy grandes, desde la red interna hacia Internet.

Los datos presentados muestran que la presencia de aplicaciones que generan tráfico a ráfagas en la red interna, podrían producir pérdida de paquetes, la cual podría aumentar si se incrementa la velocidad de la red interna (y se mantiene la tasa de salida al mismo valor). En este sentido, los capítulos \ref{cha:Buffer y rafagas} y \ref{cha:Buffer y applicaciones} presentan simulaciones con diferentes aplicaciones multimedia, velocidades de acceso y tamaños de \textit{buffer}, y en todos los casos la pérdida de paquetes es mayor para redes de $ 100 \; Mbps $ que para las de $ 10 \; Mbps $.

Además, se ha observado que el tráfico a ráfagas que generan algunas aplicaciones afectan a otros servicios que comparten el mismo enlace. Con la finalidad de mostrar el efecto de la naturaleza a ráfagas del tráfico de estas aplicaciones, se ha medido el \gls{mos} en llamadas de \gls{voip} concurrentes. Los datos muestran que dichas llamadas sólo son capaces de obtener una calidad media, fallando en alcanzar mejores resultados incluso cuando la utilización del enlace es del $ 70 \% $. Ya que la causa de este problema es la naturaleza a ráfagas de muchas aplicaciones, en estos casos las técnicas de conformado de tráfico, que permiten modificar la manera en que el tráfico es generado, se pueden considerar como una ventaja.

% bibliography
\printbibliography
\cleardoublepage

\end{document}